\documentclass[journal]{IEEEtran}

\usepackage{color}
\usepackage[utf8]{inputenc}
\usepackage{times}
\usepackage{amsmath}
\usepackage{amssymb}
\usepackage{amsthm}
\usepackage{mathrsfs}
\usepackage{graphicx}
\usepackage{enumerate}
\usepackage[compress]{cite}
\usepackage{float}
\usepackage[acronym]{glossaries}
\usepackage{multirow}
\usepackage{algorithm}
\usepackage{algorithmic}
\usepackage{textcomp}
\usepackage{etoolbox}
\usepackage{xcolor}
\usepackage{amsfonts}
\usepackage{mathtools}

\usepackage[caption=false]{subfig}

\newacronym{3GPP}{3GPP}{third generation partnership project}
\newacronym{5G}{5G}{fifth generation}
\newacronym{6G}{6G}{sixth generation}
\newacronym{AP}{AP}{access point}
\newacronym{ASW}{ASW}{access point switching}
\newacronym{AWGN}{AWGN}{additive white Gaussian noise}
\newacronym[longplural={angles of arrival}]{AoA}{AoA}{angle of arrival}
\newacronym[longplural={angles of departure}]{AoD}{AoD}{angle of departure}
\newacronym{B5G}{B5G}{beyond fifth generation}
\newacronym{BCD}{BCD}{block coordinate descend}
\newacronym{BRPA}{BRPA}{balanced random pilot assignment}
\newacronym{BS}{BS}{base station}
\newacronym{CAP}{CAP}{compress-after-precoding}
\newacronym{CB}{CB}{conjugate beamforming}
\newacronym{CBP}{CBP}{compress-before-precoding}
\newacronym{CCDF}{CCDF}{complementary cumulative distribution function}
\newacronym{CDF}{CDF}{cumulative distribution function}
\newacronym{CF-mMIMO}{CF-mMIMO}{cell-free massive multiple input multiple output}
\newacronym{CGA}{CGA}{constrained genetic algorithm}
\newacronym{GA}{GA}{genetic algorithm}
\newacronym{PDGA}{PDGA}{Pareto-driven genetic algorithm}
\newacronym{PPO}{PPO}{proximal policy optimization}
\newacronym{DRL}{DRL}{deep reinforcement learning}
\newacronym{DQL}{DQL}{deep Q-learning}
\newacronym{DDQN}{DDQN}{double deep Q-network}
\newacronym{DDPG}{DDPG}{deep deterministic policy gradient}
\newacronym{CoMP}{CoMP}{coordinated multipoint}
\newacronym{CPA}{CPA}{cluster-based pilot assignment}
\newacronym{CPU}{CPU}{central processing unit}
\newacronym{C-RAN}{C-RAN}{cloud radio access network}
\newacronym{CSI}{CSI}{channel state information}
\newacronym{ChiS}{ChiS}{chi-square test}
\newacronym{ChiS-ASO}{ChiS-ASO}{chiS-based \gls{ASW}}
\newacronym{DCPA}{DCPA}{dissimilarity cluster-based pilot assignment}
\newacronym{DL}{DL}{downlink}
\newacronym{EE}{EE}{energy efficiency}
\newacronym{GoF}{GoF}{goodness-of-fit}
\newacronym{GOPA}{GOPA}{globally optimal power allocation}
\newacronym{KS}{KS}{Kolmogorov-Smirnov}
\newacronym{KS-ASO}{KS-ASO}{KS-based \gls{ASW}}
\newacronym{LLH}{LLH}{Log-likelihood test}
\newacronym{LLH-ASO}{LLH-ASO}{LLH-based \gls{ASW}}
\newacronym{LOS}{LOS}{line-of-sight}
\newacronym{LS}{LS}{least-squares}
\newacronym{LSE}{LSE}{logarithmic statistical energy test}
\newacronym{LSE-ASO}{LSE-ASO}{logarithmic statistical energy \gls{ASW}}
\newacronym{MD-ASO}{MD-ASO}{mixture discrepancy-based greedy \gls{ASW}}
\newacronym{MDP}{MDP}{Markov decision process}
\newacronym{MEC}{MEC}{mobile edge computing}
\newacronym{MIMO}{MIMO}{multiple-input multiple-output}
\newacronym{M-MIMO}{M-MIMO}{massive MIMO}
\newacronym{MMSE}{MMSE}{minimum mean square error}
\newacronym{mmWave}{mmWave}{millimeter wave}
\newacronym{MR}{MR}{maximum-ratio}
\newacronym{MRC}{MRC}{maximal ratio combining}
\newacronym{MS}{MS}{mobile station}
\newacronym{ML}{ML}{machine learning}
\newacronym{MSE}{MSE}{mean square error}
\newacronym{MPL-ASO}{MPL-ASO}{minimum propagation losses-aware \gls{ASW}}
\newacronym{MU-MIMO}{MU-MIMO}{multiuser-MIMO}
\newacronym{NCB}{NCB}{normalized conjugate beamforming}
\newacronym{NMRC}{NMRC}{normalized maximal ratio combiner}
\newacronym{NN-ASO}{NN-ASO}{nearest neighbour-based \gls{ASW}}
\newacronym{NOPA}{NOPA}{normalized optimal power allocation}
\newacronym{NLOS}{NLOS}{non-line-of-sight}
\newacronym{OG-ASO}{OG-ASO}{optimal energy efficiency-based greedy \gls{ASW}}
\newacronym{PDF}{PDF}{probability density function}
\newacronym{QoS}{QoS}{quality of service}
\newacronym{rms}{rms}{root mean square}
\newacronym{RL}{RL}{reinforcement learning}
\newacronym{RHS}{RHS}{right hand side}
\newacronym{RPA}{RPA}{random pilot assignment}
\newacronym{RS-ASO}{RS-ASO}{random selection \gls{ASW}}
\newacronym{RRM}{RRM}{radio resource management}
\newacronym{RF}{RF}{radio frequency}
\newacronym{SE}{SE}{spectral efficiency}
\newacronym{SINR}{SINR}{signal-to-interference-plus-noise ratio}
\newacronym{SNR}{SNR}{signal-to-noise ratio}
\newacronym{SOPA}{SOPA}{sequential optimal power allocation}
\newacronym{SR-ASO}{SR-ASO}{spatial regularity-based greedy \gls{ASW}}
\newacronym{TDD}{TDD}{time division duplexing}
\newacronym{UDN}{UDN}{ultra dense network}
\newacronym{ULA}{ULA}{uniform linear array}
\newacronym{UPA}{UPA}{uniform planar array}
\newacronym{UL}{UL}{uplink}
\newacronym{ZF}{ZF}{zero-forcing}

\newacronym{1g}{1G}{first-generation}
\newacronym{4g}{4G}{fourth-generation}
\newacronym{5g}{5G}{fifth-generation}
\newacronym{mimo}{MIMO}{multiple-input multiple-output}
\newacronym{ris}{RIS}{reconfigurable inteligent surface}
\newacronym{siso}{SISO}{single-input single-output}
\newacronym{mmimo}{mMIMO}{massive multiple-input multiple-output}
\newacronym{cfmmimo}{CF-mMIMO}{cell-free massive multiple-input multiple-output}
\newacronym{sumimo}{SU-MIMO}{single user MIMO}
\newacronym{mumimo}{MU-MIMO}{multi user MIMO}
\newacronym{embms}{eMBMS}{evolved Multimedia Broadcast and Multicast Service}
\newacronym{sca}{SCA}{successive convex approximation}
\newacronym{sinr}{SINR}{signal-to-interference-plus-noise ratio}
\newacronym{ula}{ULA}{uniform linear array}
\newacronym{uaf}{UatF}{\emph{use-and-then-forget}}
\newacronym{mcs}{MCS}{modulation and coding scheme}
\newacronym{dcc}{DCC}{dynamic cooperation clustering}
\newacronym{mrt}{MRT}{maximum ratio transmission}
\newacronym{pmmse}{P-MMSE}{partial MMSE}
\newacronym{pzf}{P-ZF}{partial ZF}
\newacronym{zf}{ZF}{zero-forcing}
\newacronym{mr}{MR}{maximum ratio}
\newacronym{se}{SE}{spectral efficiency}
\newacronym{ee}{EE}{energy efficiency}
\newacronym{ap}{AP}{access point}
\newacronym{cpu}{CPU}{central processing unit}
\newacronym{uc}{UC}{user centric}
\newacronym{sse}{SumSE}{sum spectral efficiency}
\newacronym{mise}{MinSE}{minimum spectral efficiency}
\newacronym{asd}{ASD}{angular standard deviation}
\newacronym{adr}{ADR}{aggregated data rate}
\newacronym{embb}{eMBB}{enhanced mobile broadband}
\newacronym{mmtc}{mMTC}{massive machine type communications}
\newacronym{urllc}{URLLC}{ultra reliable low latency communications}
\newacronym{csi}{CSI}{channel state information}
\newacronym{pmi}{PMI}{precoding matrix indicator}
\newacronym{ri}{RI}{rank indicator}
\newacronym{csi-rs}{CSI-RS}{CSI-reference signal}
\newacronym{cri}{CRI}{CSI-RS resource indicator}
\newacronym{bs}{BS}{base station}
\newacronym{re}{RE}{resource element}
\newacronym{mmwave}{mmWave}{millimeter-wave}
\newacronym{umwave}{$\mu$mWaves}{micrometer waves}
\newacronym{rnn}{RNN}{recurrent neural network}
\newacronym{cnn}{CNN}{convolutional neural network}
\newacronym{ngmn}{NGMN}{next-generation mobile network}
\newacronym{lte}{LTE}{Long Term Evolution}
\newacronym{lte-a}{LTE-A}{Long Term Evolution Advanced}
\newacronym{5gnr}{5G NR}{5G New Radio}
\newacronym{mm}{MM}{mixed mode}
\newacronym{cdf}{CDF}{cumulative distribution function}
\newacronym{phy}{PHY}{physical}
\newacronym{mac}{MAC}{medium access control}
\newacronym{3gpp}{3GPP}{3rd Generation Partnership Project}
\newacronym{fdd}{FDD}{frequency division duplexing}
\newacronym{tdd}{TDD}{time division duplexing}
\newacronym{ofdm}{OFDM}{orthogonal frequency division multiplexing}
\newacronym{ss}{SS}{synchronization signal} 
\newacronym{pss}{PSS}{primary synchronization signal} 
\newacronym{sss}{SSS}{secondary synchronization signal} 
\newacronym{pbch}{PBCH}{physical broadcast channel} 
\newacronym{dmrs}{DMRS}{demodulation reference signal} 
\newacronym{gnb}{gNB}{next generation nodeB} 
\newacronym{rsrp}{RSRP}{reference signal received power} 
\newacronym{rrm}{RRM}{radio resource management} 
\newacronym{srs}{SRS}{sounding reference signal} 
\newacronym{ran}{RAN}{radio access network} 
\newacronym{nn}{NN}{neural network} 
\newacronym{ue}{UE}{user equipment} 
\newacronym{awgn}{AWGN}{additive white Gaussian noise} 
\newacronym{epa}{EPA}{Extended Pedestrian A model}
\newacronym{eva}{EVA}{Extended Vehicular A model}
\newacronym{etu}{ETU}{Extended Typical Urban model}
\newacronym{tdl}{TDL}{tapped delay line}
\newacronym{cdl}{CDL}{clustered delay line}
\newacronym{uma}{UMa}{urban macro-cell}
\newacronym{isd}{ISD}{inter-site distance}
\newacronym{nlos}{NLOS}{non-line of sight}
\newacronym{los}{LOS}{line of sight}
\newacronym{o2o}{O2O}{outdoor-to-outdoor}
\newacronym{o2i}{O2I}{outdoor-to-indoor}
\newacronym{ul}{UL}{uplink}
\newacronym{dl}{DL}{downlink}
\newacronym{ls}{LS}{least squares}
\newacronym{mmse}{MMSE}{minimum mean square error}
\newacronym{snr}{SNR}{signal-to-noise ratio}
\newacronym{mse}{MSE}{mean square error}
\newacronym{nr}{NR}{New Radio}
\newacronym{prb}{PRB}{physical resource block}
\newacronym{scs}{SCS}{subcarrier spacing}
\newacronym{bler}{BLER}{block error rate}
\newacronym{smmmra}{SMMMRA}{subgroup multicast \gls{mamimo} resource allocation}
\newacronym{mmf}{MMF}{max-min fairness}
\newacronym{smmu}{SMMU}{subgroups of multicast \gls{mamimo} users}
\newacronym{gsmma}{GSMMA}{greedy subgroup multicast \gls{mamimo} algorithm}
\newacronym{apa}{APA}{adaptive power allocation}
\newacronym{ms}{MS}{mobile station}
\newacronym{cb}{CB}{conjugate beamforming}
\newacronym{ncb}{NCB}{normalized CB}
\newacronym{ecb}{ECB}{enhanced normalized CB}
\newacronym{noma}{NOMA}{non-orthogonal multiple access}
\newacronym{xr}{XR}{extended reality}
\newacronym{adc}{ADC}{analog-digital converter}
\newacronym{dac}{DAC}{digital-analog converter}

\newcommand{\bs}{\boldsymbol}

\DeclareMathOperator{\blockdiag}{blockdiag}

\DeclareMathOperator{\col}{col}
\newcommand{\herm}{^\mathsf{H}}
\newcommand{\trans}{^\mathsf{T}}

\def\BibTeX{{\rm B\kern-.05em{\sc i\kern-.025em b}\kern-.08em
    T\kern-.1667em\lower.7ex\hbox{E}\kern-.125emX}}

\makeatletter
\renewcommand{\thesubsection}{\thesection-\Alph{subsection}}
\makeatother

\begin{document}

\title{Evolutionary AP Switch ON/OFF Techniques for Energy-efficient Cell-free Massive MIMO Networks}

\author{
Jan~Garc\'ia-Morales,
Alejandro~de~la~Fuente,
David~Gualda,
Leopoldo~Carro-Calvo,
Felip~Riera-Palou,~\IEEEmembership{Senior Member, IEEE},
and Guillem~Femenias,~\IEEEmembership{Senior Member, IEEE}
\thanks{Jan García-Morales, Alejandro de la Fuente, and David Gualda are with Rey Juan Carlos University (URJC), Fuenlabrada, Madrid 28942, Spain.}
\thanks{Leopoldo Carro-Calvo is with the Institute of Data, Complex Networks and Cybersecurity Sciences (DCNC Sciences), Rey Juan Carlos University, Madrid 28028, Spain.}
\thanks{Felip Riera-Palou and Guillem Femenias are with the SECOM Group and the Artificial Intelligence Research Institute (IAIB), University of the Balearic Islands (UIB), Palma 07122, Spain.}
\thanks{Corresponding author: Jan García-Morales (e-mail: jan.garcia@urjc.es).}
\thanks{
Author Accepted Manuscript of the article published in IEEE Open Journal of the Communications Society.
The final published version is available at
https://doi.org/10.1109/OJCOMS.2026.3717118.
Copyright © 2026 IEEE.
}
\thanks{This work was supported by grants SOFIA-WIND (PID2023-147305OB-C33), brAIn5G (PID2024-161515OA-I00), POLIGRAPH (PID2022-136887NB-I00), and AI-XCAST6G (F1263).}
}

\maketitle

\begin{abstract}
\Gls{CF-mMIMO} is an emerging technology for next-generation wireless systems, where dynamically adapting the set of active \glspl{AP} is crucial to balance \gls{QoS} requirements and network energy consumption under highly time-varying and spatially non-uniform traffic loads.
Existing \gls{AP} ON/OFF mechanisms—typically based on worst‑case dimensioning or greedy heuristics—explore the combinatorial activation space inadequately, leading to suboptimal energy‑efficiency outcomes.
This paper introduces two evolutionary \gls{AP}‑selection strategies tailored to \gls{CF-mMIMO} networks.
The first, a \gls{CGA}, identifies the near‑optimal subset of active \glspl{AP} for any fixed activation cardinality, while an outer search determines the globally optimal operating point.
The second, a \gls{PDGA}, jointly optimizes spectral and energy efficiency by evolving a Pareto front over all feasible activation patterns.
A detailed computational‑complexity analysis is provided for both techniques.
Simulations conducted under realistic spatially inhomogeneous traffic and considering both \gls{CB} and \gls{MMSE} processing confirm consistent performance gains.
The proposed methods consistently outperform state-of-the-art greedy benchmarks, delivering noticeable improvements in energy efficiency for both \gls{CB} and \gls{MMSE} schemes, while simultaneously enhancing the energy–spectral efficiency tradeoff, which is typically difficult to improve without incurring penalties elsewhere.
These results highlight the strong potential of evolutionary optimization as a powerful and reliable approach for energy-efficient \Gls{CF-mMIMO} deployments.
\end{abstract}


\glsresetall

\section{INTRODUCTION}
\IEEEPARstart{C}{ell}-free massive multiple input multiple output (\glsentryshort{CF-mMIMO}) has emerged as a leading architecture for next-generation wireless systems, enabling uniformly high service quality through the coordinated transmission of a large number of distributed \glspl{AP} controlled by a \gls{CPU}.
\glsunset{CF-mMIMO}
By exploiting favorable propagation and channel hardening, \gls{CF-mMIMO} provides strong macro-diversity and mitigates cell-edge effects, ensuring that \glspl{MS} experience nearly uniform performance across the service area~\cite{2017Ngo,2019Interdonato,2019Femenias,2024Ngo}. In practical deployments, a large number of geographically scattered \glspl{AP} jointly serve a comparatively smaller number of \glspl{MS} over shared time–frequency resources, enabling improved throughput, fairness, and coverage uniformity.

Despite these advantages, the massive densification of radio-access infrastructure and the fully cooperative nature of \gls{CF-mMIMO} impose significant energy consumption. Projections indicate that the information and communication technology (ICT) sector may account for up to 23\% of global carbon emissions and more than 50\% of global electricity usage by 2030~\cite{Andrae15}, making energy-efficient (“green”) wireless communications a critical research priority~\cite{leong2024green}.
In this context, dynamically adapting the set of active \glspl{AP} to spatio-temporal variations in network load is a highly effective strategy to reduce energy consumption while maintaining \gls{QoS} guarantees. Techniques such as sleep-mode control, \gls{MS} association, load balancing, and cell zooming have been investigated~\cite{Wu15,Han16,Gandotra17,Tabassum14,Jia15,Xu17,Jiang18}, but many rely on the unrealistic assumption of spatially homogeneous \gls{MS} distributions. Real mobile networks exhibit strong spatial inhomogeneity, with irregular hotspots driven by device density, mobility, and application diversity~\cite{lee2014spatial}. These factors motivate the design of \gls{AP} activation strategies that explicitly account for heterogeneous traffic patterns.

The problem addressed in this paper is the following: given a large-scale \gls{CF-mMIMO} network with spatially inhomogeneous traffic, determine the subset of active \glspl{AP} that optimizes \gls{EE} while satisfying \gls{QoS} constraints. This optimization task is combinatorial and NP-hard, as the number of feasible ON/OFF configurations grows exponentially with the number of \glspl{AP}. For instance, a deployment with $L = 100$ \glspl{AP} leads to $2^{100}$ possible activation patterns, making exhaustive search entirely infeasible even under idealized assumptions. The impact of activation decisions is further amplified by the choice of processing scheme: under \gls{CB}, inter-\gls{MS} interference is not actively suppressed but this processing entails a very low computational burden, whereas \gls{MMSE} processing can exploit richer \gls{AP} activation patterns thanks to its interference-mitigation capabilities at the cost of a large computational complexity.
Efficiently navigating this vast solution space under heterogeneous traffic conditions remains an open challenge and serves as the core motivation for this work.
Beyond the combinatorial complexity of \gls{AP} activation, a key challenge lies in balancing competing performance objectives, notably \gls{EE} and \gls{SE}. In practical \gls{CF-mMIMO} deployments, operators are often interested in understanding the trade-offs between these metrics rather than identifying a single operating point. This perspective naturally motivates a multi-objective optimization formulation, which enables the explicit characterization of the achievable \gls{EE}–\gls{SE} trade-off and supports more informed and flexible network configuration decisions.
To the best of our knowledge, no prior work has investigated evolutionary multi-objective optimization for the \gls{AP} activation problem in large-scale \gls{CF-mMIMO} networks under realistic spatially heterogeneous traffic.

\subsection{RELATED WORK}
Classical \gls{ASW} strategies for \gls{CF-mMIMO} predominantly rely on greedy and rule-based heuristics owing to their low computational complexity.
Early works focused on homogeneous \gls{MS} distributions~\cite{femenias2020access}, while \cite{garcia2020energy} incorporated heterogeneous layouts through \gls{GoF} metrics to align \gls{AP} density with traffic intensity.
Extensions include heuristics based on effective channel gains~\cite{jung2021performance,di2025digital}, two-timescale schemes leveraging \gls{MS} mobility~\cite{riera2021two}, and adaptations to macro-cellular 5G scenarios~\cite{riera2025sleep}.
More recent formulations consider joint \gls{EE}–\gls{SE} optimization~\cite{jayaweera2024minimizing}.
Although computationally efficient, greedy and rule-based heuristics explore only a limited portion of the combinatorial space, lack global coordination, and cannot recover from early suboptimal decisions—limitations that become critical under spatially heterogeneous traffic.

\begin{table*}[t!]
\renewcommand{\arraystretch}{0.9}
\caption{\small Comparative overview of representative RL/DRL-based and alternative evolutionary optimization-based AP ON/OFF approaches for CF-mMIMO networks}
\label{tab:rl_comparison}
\centering
\begin{tabular}{p{0.8cm} p{1.8cm} p{2.0cm} p{2.2cm} p{2.4cm} p{1.8cm} p{3cm}}
\hline
\bfseries ~Ref. &
\bfseries Approach &
\bfseries MS distribution &
\bfseries Evaluation scale &
\bfseries Temporal scale &
\bfseries Main focus &
\bfseries Technical considerations\\
\hline
\footnotesize {\cite{mendoza2021deep}} &
\footnotesize {Centralized DDQN} &
\footnotesize {Uniform} &
\footnotesize {Small (8 APs)} &
\footnotesize {Long-term, \text{ }\text{ }\text{ }\text{ }\text{ }\text{ } Per-realization CSI} &
\footnotesize {QoS \text{ }\text{ }\text{ }\text{ }\text{ }\text{ }\text{ }\text{ }\text{ } optimization} &
\footnotesize {Small-scale evaluation, no EE optimization, only CB} \\
\hline
\footnotesize {\cite{suh2023drl}} &
\footnotesize {Actor--Critic DDPG} &
\footnotesize {Uniform} &
\footnotesize {Large (120 APs)} &
\footnotesize {Long-term, \text{ }\text{ }\text{ }\text{ }\text{ }\text{ } Per-realization CSI} &
\footnotesize {Power \text{ }\text{ }\text{ }\text{ }\text{ }\text{ }\text{ } minimization} &
\footnotesize {No EE optimization, \text{ }\text{ }\text{ }\text{ } only MR} \\
\hline
\footnotesize {\cite{sun2023multi}} &
\footnotesize {Multi-agent DRL} &
\footnotesize {Uniform} &
\footnotesize {Moderate (50 APs)} &
\footnotesize {Long-term, \text{ }\text{ }\text{ }\text{ }\text{ }\text{ } Per-realization CSI} &
\footnotesize {Power \text{ }\text{ }\text{ }\text{ }\text{ }\text{ }\text{ } minimization} &
\footnotesize {No EE optimization, \text{ }\text{ }\text{ }\text{ } only CB} \\
\hline
\footnotesize {\cite{li2024energy}} &
\footnotesize {PPO-based} &
\footnotesize {Uniform} &
\footnotesize {Moderate (50 APs)} &
\footnotesize {Long-term, \text{ }\text{ }\text{ }\text{ }\text{ }\text{ } Per-realization CSI} &
\footnotesize {EE \text{ }\text{ }\text{ }\text{ }\text{ }\text{ }\text{ }\text{ }\text{ }\text{ } maximization} &
\footnotesize {Shallow search, \text{ }\text{ }\text{ }\text{ }\text{ }\text{ }\text{ } LP-MMSE} \\
\hline
\footnotesize {\cite{garcia2026energy}} &
\footnotesize {PPO-based} &
\footnotesize {Uniform} &
\footnotesize {Large (96 APs)} &
\footnotesize {Long-term, \text{ }\text{ }\text{ }\text{ }\text{ }\text{ } Per-realization CSI} &
\footnotesize {EE \text{ }\text{ }\text{ }\text{ }\text{ }\text{ }\text{ }\text{ }\text{ }\text{ }  maximization} &
\footnotesize {Shallow search, \text{ }\text{ }\text{ }\text{ }\text{ }\text{ }\text{ } LP-MMSE} \\
\hline
\footnotesize {This work} &
\footnotesize {GA-based} &
\footnotesize {Heterogeneous} &
\footnotesize {Large (100 APs)} &
\footnotesize {Very long-term, \text{ }\text{ }\text{ }\text{ }\text{ } Per-distribution CSI} &
\footnotesize {EE \text{ }\text{ }\text{ }\text{ }\text{ }\text{ }\text{ }\text{ }\text{ }\text{ } maximization} &
\footnotesize {Deep search, CB and MMSE} \\
\hline
\end{tabular}
\end{table*}

Approximate dynamic programming (ADP) and \gls{RL}/\gls{DRL} have recently been explored for \gls{AP} activation~\cite{mendoza2021deep,suh2023drl,sun2023multi,li2024energy,garcia2026energy}.
Centralized \gls{DRL} approaches, such as double deep Q-network (DDQN)~\cite{mendoza2021deep}, have demonstrated promising results in very small-scale networks (e.g., 8 \glspl{AP}), where the number of \glspl{AP} and the corresponding action space remain limited. However, when extending these methods to realistic \gls{CF-mMIMO} deployments with tens or hundreds of \glspl{AP} (e.g., $100$ \glspl{AP}, $2^{100}$ configurations), the exponential growth of the combinatorial action space leads to significant training and inference challenges.
To ensure affordability in large-scale networks, \cite{suh2023drl} restricts the processing to a decentralized low-complexity \gls{MR} combining, significantly reducing \gls{CPU} load and backhaul signaling. However, incorporating delay constraints into the reward function inherently couples power consumption and latency, hindering an explicit optimization of \gls{EE} as a standalone objective.
Multi-agent \gls{DRL} approaches~\cite{sun2023multi} address dimensionality issues through decentralized decision-making, but rely exclusively on downlink \gls{CB}, whereby inter-\gls{MS} interference is not actively suppressed, thus resulting in poor spectral efficiencies. Moreover, their optimization objective is cast as power minimization under rate constraints, rather than as an explicit maximization of \gls{EE}.
\Gls{PPO}-based, centrally trained \gls{DRL} approaches~\cite{li2024energy,garcia2026energy} reduce the dimensionality of the action space by directly generating \gls{AP} activation decisions.
These works adopt large-scale fading decoding combined with local partial minimum mean-squared error (LP-MMSE) processing.
Nevertheless, although improved \gls{EE} performance have been reported in controlled scenarios, these policy-gradient methods operate under partial observability by using a shallow search and rely on realization-dependent inputs, such as \gls{MR} positions or average \gls{CSI}.
While \gls{RL}/\gls{DRL}-based techniques are reviewed for completeness and represent an active line of research, their use as baselines for the large-scale \gls{CF-mMIMO} problem considered here remains challenging from an implementation perspective.
\gls{RL}/\gls{DRL} methods aim to learn policies that map compact state observations to \gls{AP} activation decisions through offline training, whereas the problem considered here is formulated as the direct optimization of \gls{AP} ON/OFF patterns for given spatial \gls{MS} density distributions, enabling deployment-agnostic designs without prior training. 
A fair comparison with \gls{RL}/\gls{DRL} would require training over diverse distributions, load conditions, and channel realizations, significantly increasing the number of physical-layer evaluations.
While each \gls{MS} distribution can be independently optimized and explicitly stored in the considered framework, \gls{RL}/\gls{DRL} must incorporate all scenarios within one or more training processes based on extensive environment interactions, where each interaction involves \gls{MS} position tracking, action selection, physical-layer evaluation, and \gls{SE}/\gls{EE} reward computation. Thus, computational cost is not eliminated but shifted and amplified during training, and achieving comparable generality requires substantially more evaluations.
Moreover, \gls{RL}/\gls{DRL} performance critically depends on the representativeness of the training set: biases in spatial \gls{MS} distributions, traffic loads, or network geometries can degrade generalization, whereas increasing diversity raises training complexity, interaction requirements, and potential stability issues. Meaningful comparisons therefore require carefully designed training environments, validation of policy stability, and evaluation on unseen spatial \gls{MS} distributions.
Methodologically, explicit optimization approaches generate configurations that can be stored and incrementally extended as new traffic patterns emerge, without affecting prior solutions.
This distinction is particularly relevant in practical \gls{CF-mMIMO} deployments, where fixed \gls{AP} locations and statistically characterizable traffic distributions enable reusable mappings between \gls{MS} density distributions and \gls{AP} activation patterns, allowing real-time operation via low-cost lookup tables.
\gls{RL}/\gls{DRL} policies trained on limited or simplified traffic distributions (e.g., uniform \gls{MS} deployments) are not fully representative, while training over sufficiently broad scenarios would require dedicated large-scale studies focused on generalization and transferability.
Finally, \gls{RL}/\gls{DRL} methods often rely on high-dimensional, realization-dependent inputs (e.g., instantaneous \gls{MS} locations or channel states), which yield policies tied to specific operating conditions and complicate large-scale real-time implementation.
Although such approaches can reduce decision complexity at inference time, they do not eliminate the cost of physical-layer evaluations, which is instead incurred during training.
Table~\ref{tab:rl_comparison} summarizes a comparison between representative \gls{RL}/\gls{DRL}-based and evolutionary approaches for \gls{AP} ON/OFF optimization in \gls{CF-mMIMO} networks, highlighting their evaluation scale, temporal scale of operation, and key technical considerations.

\subsection{CONTRIBUTIONS}
This paper proposes evolutionary optimization strategies for \gls{AP} activation problem in \gls{CF-mMIMO} networks (categorized as \gls{GA}-based approaches in Table~\ref{tab:rl_comparison}) and evaluates their performance under heterogeneous spatial traffic distributions. 
Unlike greedy heuristics in~\cite{femenias2020access,garcia2020energy}, adopted as benchmarks, the proposed \gls{GA}-based approaches effectively explore the global combinatorial space and provide notable improvements in \gls{EE} under both \gls{CB} and \gls{MMSE} processing.
The proposed strategies operate at very large temporal scales, exploiting the slow and often cyclical evolution of traffic patterns.
The developed framework avoids prior training and enables deployment-agnostic operation.
The resulting solutions can be precomputed offline and implemented in real time via lookup tables.
The main contributions of this work are summarized as follows:

\begin{itemize}
\item \textbf{Evolutionary optimization framework for \gls{AP} activation:}  
An evolutionary search framework is developed to explore the global combinatorial space of \gls{AP} ON/OFF activation patterns, enabling the identification of high-quality activation strategies beyond the limitations of greedy methods.

\item \textbf{Constrained genetic algorithm (CGA):}  
A constrained genetic algorithm is proposed to determine near-optimal \gls{AP} activation subsets for a fixed number of active \glspl{AP}. By combining this constrained search with an external search over the activation cardinality, the method approximates a globally competitive ON/OFF activation configuration.

\item \textbf{Pareto-driven genetic algorithm (PDGA):} 
A multi-objective genetic algorithm is developed to jointly optimize \gls{EE} and \gls{SE} by constructing the Pareto front over the feasible AP activation space.
Beyond optimization, the framework explicitly characterizes the \gls{EE}–\gls{SE} trade-off in large-scale \gls{CF-mMIMO} networks, enabling a decision-oriented interpretation of the \gls{AP} ON/OFF problem. 
This allows network operators to select operating points according to system-level requirements, such as throughput or \gls{QoS} constraints.

\item \textbf{Computational-complexity analysis:}  
A detailed complexity analysis is presented to evaluate the affordability and computational cost of the proposed evolutionary strategies and to compare them with classical greedy-search baselines.

\item \textbf{Performance evaluation and design insights:}  
Simulation results under heterogeneous spatial traffic distributions demonstrate that evolutionary optimization consistently improves energy efficiency and provide insight into the relationship between the optimal number of active \glspl{AP}, the spatial traffic distribution, and the processing scheme.
\end{itemize}
We emphasize that the contribution of this work lies in the development and analysis of evolutionary optimization strategies tailored to the large-scale \gls{AP} activation problem, as well as in the characterization of the \gls{EE} and \gls{SE} trade-off in \gls{CF-mMIMO} networks.
Overall, the proposed contributions provide not only improved energy efficiency performance, but also bring new insights into the interplay between \gls{AP} activation, traffic heterogeneity, and processing schemes, while enabling practical operating-point selection in \gls{CF-mMIMO} networks.

\subsection{ORGANIZATION OF THE PAPER}
The paper is organized as follows. 
Section~\ref{sec:System_model} presents the \gls{CF-mMIMO} system model, including channel estimation, downlink transmission, and spatial \gls{MS} distribution. Section~\ref{sec:Performance_metrics} introduces the performance metrics employed to evaluate \gls{AP} activation schemes. Sections~\ref{sec:GA_ASW} and~\ref{sec:Gready_ASW} describe the proposed \gls{GA}-based and benchmark \gls{GoF}-based methods, respectively. Section~\ref{sec:numerical_results} reports numerical results, and Section~\ref{sec:Conclusion} concludes the paper.

\section{SYSTEM MODEL}
\label{sec:System_model}

\begin{figure}[t]
    \centering
    \includegraphics[width=7.8cm]{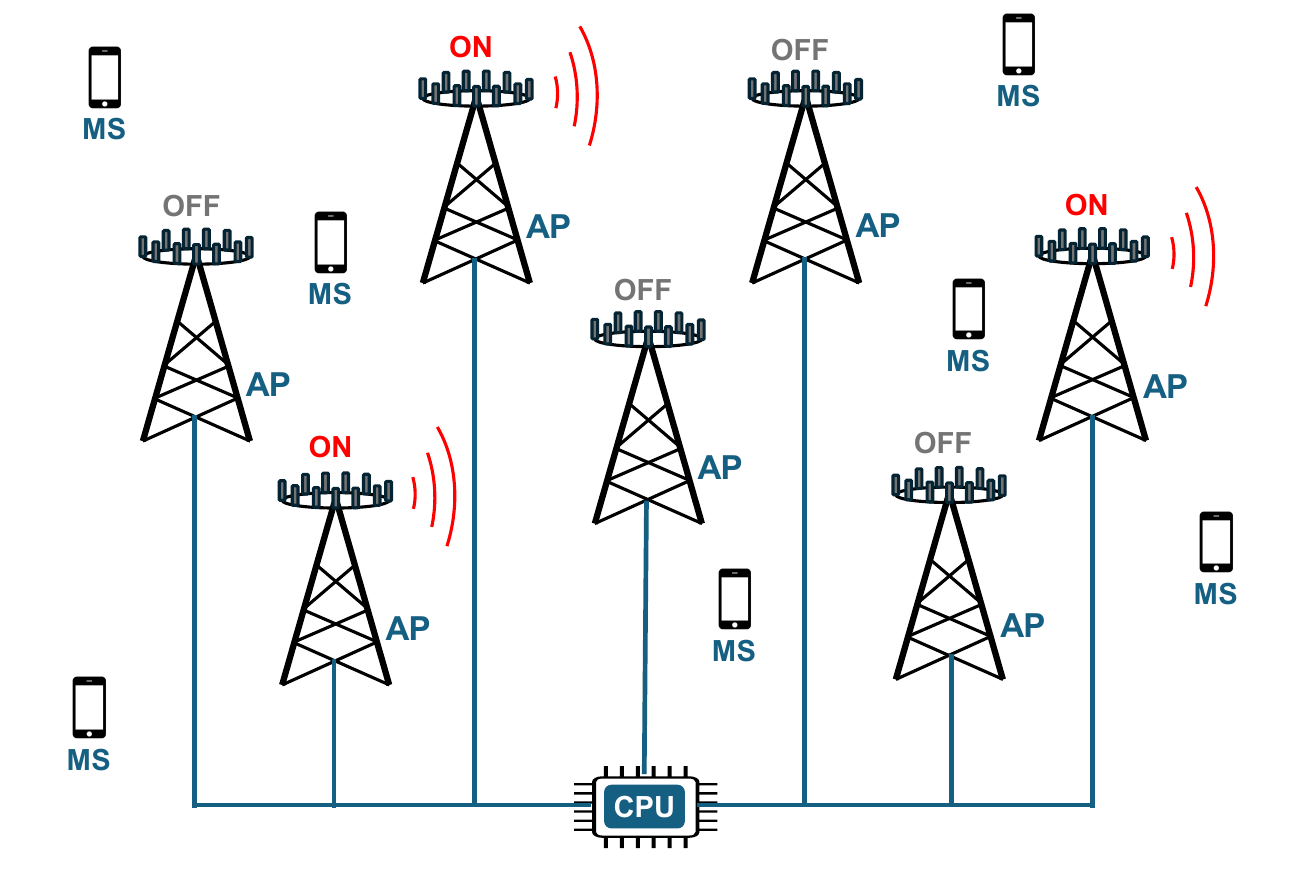}
    \caption{\Gls{CF-mMIMO} network where $L_A$ active \glspl{AP} (out of $L$ available \glspl{AP}) provide service to $K$ \glspl{MS}.}
    \label{fig:system_model}
\end{figure}

A \gls{CF-mMIMO} architecture operating under \gls{TDD} is examined, in which a total of $L$ \glspl{AP}, each equipped with $N$ antennas, are interconnected with a \gls{CPU} via ideal fronthaul links \cite{2017Ngo}.
These \glspl{AP} are scattered throughout the service region and jointly serve $K$ single antenna \glspl{MS} using the same time-frequency resources, as depicted in Fig.~\ref{fig:system_model}.

Let $\mathcal{K} = \{1, \ldots, K\}$ and $\mathcal{L} = \{1, \ldots, L\}$ represent the sets of \glspl{MS} and \glspl{AP}, respectively. Each \gls{AP} can dynamically switch between active and sleep states, denoted by the disjoint subsets $\mathcal{L}^A$ and $\mathcal{L}^S$, such that $\mathcal{L} = \mathcal{L}^A \cup \mathcal{L}^S$ and $\mathcal{L}^A \cap \mathcal{L}^S = \emptyset$. The cardinalities of these subsets are $L_A$ and $L_S$, respectively.

The coordination of data exchange is governed by the \gls{CPU}, which orchestrates a half-duplex \gls{TDD} protocol.
Each transmission cycle is segmented into distinct phases: uplink training phase, and uplink and downlink payload data transmission phases.
During the uplink training phase, all \glspl{MS} transmit predefined sequences, enabling active \glspl{AP} to estimate the channel responses. These estimates are subsequently used to design the combiners and precoders used in the uplink and downlink payload data transmission phases, respectively. The combined duration/bandwidth of the training, uplink and downlink payload data transmission phases, denoted as $\tau_p$, $\tau_u$ and $\tau_d$, respectively, should not exceed the coherence time/bandwidth of the channel, denoted as $\tau_c$, that is, $\tau_p+\tau_d+\tau_u\leq \tau_c$, with all these intervals specified in samples (or channel uses) on a time-frequency grid.
For the sake of simplicity, this study focuses on the downlink; therefore, without loss of generality, the size of the uplink payload data transmission phase is set to $\tau_u=0$.

\subsection{CHANNEL MODEL}
We adopt a standard block-fading model in which the wireless channel remains constant and frequency-flat over a coherence block, and evolves independently across different blocks. The channel vector between \gls{AP} $l$ and \gls{MS} $k$ is indicated by $\boldsymbol{h}_{lk} \in \mathbb{C}^N$ and is modeled as a complex circularly symmetric Gaussian random vector $\boldsymbol{h}_{lk} \!\sim\! \mathcal{CN}(\boldsymbol{0}_N,\boldsymbol{R}_{lk})$, where $\boldsymbol{R}_{lk} \in \mathbb{C}^{N \times N}$ is the corresponding positive semi-definite spatial covariance matrix, and the average channel gain holds $\beta_{lk} = \frac{1}{N} \text{tr}(\boldsymbol{R}_{lk})$.
Covariance matrices $\boldsymbol{R}_{lk}$ can be estimated over long-term channel statistics, i.e., over many coherence intervals, and are assumed to be perfectly known at both the \glspl{AP} and the \gls{CPU} \cite{bjornson2016massive}.

Due to the physical separation between \glspl{AP} and \glspl{MS}, it is reasonable to assume that the channels between different \gls{AP}-\gls{MS} pairs are statistically independent. That is, $\mathbb{E}\{\boldsymbol{h}_{l'k'} \boldsymbol{h}_{lk}\herm\} \!=\! \boldsymbol{0}_{N\times N}, \ \forall (l'k') \neq (lk)$.
The aggregate channel vector from \gls{MS} $k$ to all \glspl{AP} is defined as $\boldsymbol{h}_k\!=\!\col\left(\left\{\boldsymbol{h}_{lk}\right\}_{l\in\mathcal{L}}\right)$, which is distributed as $\boldsymbol{h}_{k} \!\sim\! \mathcal{CN}(\boldsymbol{0}_{LN},\boldsymbol{R}_{k})$, where $\boldsymbol{R}_{k} =\blockdiag(\boldsymbol{R}_{1k},\ldots,\boldsymbol{R}_{Lk})$.

\subsection{CHANNEL ESTIMATION}
In the uplink training phase, each \gls{MS} transmits a pilot sequence to facilitate channel estimation at the \glspl{AP}.
A set of ${\tau_p}$ mutually orthogonal pilot sequences are used during the uplink training phase.
Let $\boldsymbol{\psi}_k \in \mathbb{C}^{\tau_p}$ denote the pilot sequence assigned to \gls{MS} $k$, normalized such that $\left\|\boldsymbol{\psi}_k\right\|^2=1$. The $N \!\times\! \tau_p$ signal matrix received at \gls{AP} $l$ during this phase is obtained as
\begin{equation}    
    \boldsymbol{Y}_l = \sqrt{\tau_p P_p}\sum_{k \in \mathcal{K}} \boldsymbol{h}_{lk}\boldsymbol{\psi}\trans_k + \boldsymbol{N}_l,
\end{equation}
where $P_p$ is the pilot power per symbol and $\boldsymbol{N}_l$ is the \gls{AWGN} matrix with i.i.d. entries distributed as $\mathcal{CN}(0, \sigma_u^2)$.
In scenarios where the number of \glspl{MS} exceeds the number of available orthogonal pilot sequences ($K > \tau_p$), pilot reuse becomes necessary, leading to interference during channel estimation, commonly termed \emph{pilot contamination} \cite{Marzetta10,Marzetta16,Elijah16}.
To extract channel information for \gls{MS} $k$, AP $l$ correlates the received signal with the conjugate of the pilot as
\begin{equation}
\boldsymbol{y}_{lk}=\boldsymbol{Y}_l \boldsymbol{\psi}_k^* \!=\! \sqrt{\tau_pP_p}\boldsymbol{h}_{lk}\!+\!\sqrt{\tau_pP_p} \sum_{i \in \mathcal{P}_k\setminus k} \boldsymbol{h}_{li} \!+ \boldsymbol{n}_{lk},
\label{eq:y_lg}
\end{equation}
where $\mathcal{P}_k$ is the set of \glspl{MS} that have been assigned the same pilot sequence as \gls{MS} $k$, including itself, and $\boldsymbol{n}_{lk} = \boldsymbol{N}_l \boldsymbol{\psi}^{\ast}_k  \sim \mathcal{CN}(0,\sigma_u^2\mathbf{I}_N)$.

The \gls{MMSE} estimate of $\boldsymbol{h}_{lk}$ is computed as
\begin{equation}
     \hat{\boldsymbol{h}}_{lk} = \ \sqrt{\tau_p P_p}\boldsymbol{R}_{lk} \ \boldsymbol{\Gamma}_{lk}^{-1} \ \boldsymbol{y}_{lk} \, , \label{eq:h_lk_est}
\end{equation}
with
\begin{equation}
    \label{eq:gamma_lk}
     \boldsymbol{\Gamma}_{lk} =\tau_p P_p\! \sum\limits_{i \in \mathcal{P}_k} \boldsymbol{R}_{li} + \sigma_u^2\boldsymbol{I}_{N}.
\end{equation}

The estimated channel is distributed as $\hat{\boldsymbol{h}}_{lk}  \!\sim\! \mathcal{CN}(\boldsymbol{0}_N, \tau_p P_p\boldsymbol{R}_{lk} \boldsymbol{\Gamma}_{lk}^{-1} \boldsymbol{R}_{lk})$, and is uncorrelated with the estimation error defined as $\tilde{\bs{h}}_{lk}=\bs{h}_{lk}-\hat{\bs{h}}_{lk}$ and distributed as $\tilde{\boldsymbol{h}}_{lk} \!\sim\! \mathcal{CN}(\boldsymbol{0}_N, \boldsymbol{R}_{lk} \!-\! \tau_p P_p\boldsymbol{R}_{lk} \boldsymbol{\Gamma}_{lk}^{-1} \boldsymbol{R}_{lk})$.

\subsection{DOWNLINK PAYLOAD DATA TRANSMISSION}
During the downlink phase, the active \glspl{AP} in the set $\mathcal{L}^A$ deliver data to the \glspl{MS} in $\mathcal{K}$ using linear precoding techniques. We adopt the canonical \gls{CF-mMIMO} transmission model in which all active \glspl{AP} jointly serve all \glspl{MS}. The signal received by \gls{MS} $k$ is expressed as
\begin{equation}
      y_{k} \!=\!  \sum_{l\in\mathcal{L}^A}\boldsymbol{h}\herm_{lk}  \mathbf{w}_{lk}\varsigma_k \!+\! \sum_{l\in\mathcal{L}^A}\sum^K_{\substack{i=1 \\ i\neq k}} \boldsymbol{h}\herm_{lk}  \mathbf{w}_{li}\varsigma_i \!+\! n_k,   \label{eq:y_gk}
\end{equation} 
where $n_k \sim \mathcal{CN}(0, \sigma_d^2)$ denotes the receiver \gls{AWGN} at \gls{MS} $k$, $\mathbf{w}_{lk} \in \mathbb{C}^{N}$ is the precoding vector used by \gls{AP} $l$ for \gls{MS} $k$, and $\varsigma_k$ is the data symbol intended for \gls{MS} $k$, satisfying $\mathbb{E}\{|\varsigma_k|^2\} = 1$ and $\mathbb{E}\{\varsigma_k \varsigma_i^*\} = 0$ for $k \neq i$. The first term represents the desired signal, the second term accounts for multi-\gls{MS} interference, and the third term is the \gls{AWGN}.

We explore two distinct strategies for precoding and power control: a centralized scheme based on the \gls{MMSE} strategy and a distributed approach based on \gls{CB}.

\subsubsection{CENTRALIZED MMSE PRECODING}
In the centralized configuration, the \gls{CPU} has access to the channel estimates of all \glspl{MS}, allowing for the joint design of precoding vectors.
Let $\hat{\boldsymbol{h}}_k = \col(\{\hat{\boldsymbol{h}}_{lk}\}_{l\in\mathcal{L}^A}) \in \mathbb{C}^{L_AN}$ be the aggregate channel estimate for \gls{MS} $k$, and
$\mathbf{w}_k =  \col\left(\left\{\mathbf{w}_{lk}\right\}_{l\in\mathcal{L}^A}\right) \in \mathbb{C}^{L_AN}$ the corresponding precoding vector. The received signal can then be compactly written as
\begin{equation}
      y_{k} \!=\!  \boldsymbol{h}\herm_{k}  \mathbf{w}_{k}\varsigma_k \!+\! \sum^K_{\substack{i=1 \\ i\neq k}} \boldsymbol{h}\herm_{k}  \mathbf{w}_{i}\varsigma_i \!+\! n_k,   \label{eq:centralized:y_gk}
\end{equation}
where $\boldsymbol{h}_k = \col\left(\left\{\boldsymbol{h}_{lk}\right\}_{l\in\mathcal{L}^A}\right) \in \mathbb{C}^{L_AN}$ is the channel vector.
The precoding vector $\mathbf{w}_k$ is derived using the uplink-downlink duality framework \cite{2021Demir}, and is given by 
\begin{equation}
      \mathbf{w}^{\text{MMSE}}_k = \sqrt{\rho_k} \frac{\bar{\mathbf{w}}^{\text{MMSE}}_k}{\sqrt{\mathbb{E}\{\|\bar{\mathbf{w}}^{\text{MMSE}}_k\|^2\}}},
      \label{eq:precoding_vector}
\end{equation}
where $\rho_k$ is the power control coefficient applied to \gls{MS} $k$, and $\bar{\mathbf{w}}^{\text{MMSE}}_k$ is the dual uplink combining vector, defined as
\begin{equation}
           \bar{\mathbf{w}}^{\text{MMSE}}_k \!=\! {p_k}\Bigg(\sum_{i=1}^K p_i \boldsymbol{\hat{h}}_{i} \boldsymbol{(\hat{h}}_{i})\herm +  \boldsymbol{C}_i + \sigma_{\mathrm{u}}^2 \boldsymbol{I}_{L_AN} \Bigg)^{\!\!\!-1}\! \boldsymbol{\hat{h}}_k,
\end{equation}
with $\boldsymbol{C}_i$ being the covariance matrix of $\tilde{\boldsymbol{h}}_i$ and $p_k$ representing the uplink power for \gls{MS} $k$.

To regulate downlink transmit power across \glspl{MS}, we employ the fractional power control mechanism as \cite{2021Demir}
\begin{equation}
    \begin{split}
     \rho_{k} = P_{\mathrm{dl}}\frac{\bigg(\sum\limits_{l \in \mathcal{L}^A} \beta_{lk} \bigg)^\nu \omega_k^{-\kappa}} {\underset{i \in \mathcal{K}}\max \Bigg[ \bigg(\sum\limits_{l \in \mathcal{L}^A} \beta_{li} \bigg)^\nu \omega_i^{1-\kappa}\Bigg] },
     \end{split}
    \label{eq:p_g}
\end{equation}
where $P_{\mathrm{dl}}$ is the maximum transmit power per \gls{AP}, and $\nu \in [-1,1]$ adjusts the fairness-performance trade-off with negative and positive values approaching max-min and sum-rate designs, respectively.
In addition, we employ
\begin{equation}
      \omega_k = \max_{l \in \mathcal{L}^A} \frac{\mathbb{E}\{\|\bar{\mathbf{w}}^{\text{MMSE}}_{lk}\|^2\}}{\mathbb{E}\{\|\bar{\mathbf{w}}^{\text{MMSE}}_k\|^2\}},
      \label{eq:omega_k}
\end{equation}
as the peak fraction of $\rho_k$ used by any AP, which can be further shaped using an exponent $\kappa \in [0,1]$ to refine the power distribution between \glspl{MS}. 
This strategy ensures that the following power constraint is satisfied
\begin{equation}
      P_{l}^{\text{Tx}}=\sum_{k \in \mathcal{K}} \rho_k \frac{\mathbb{E}\{\|\bar{\mathbf{w}}^{\text{MMSE}}_{lk}\|^2\}}{\mathbb{E}\{\|\bar{\mathbf{w}}^{\text{MMSE}}_k\|^2\}} \leq P_{\mathrm{dl}}.
      \label{eq:power_constraint}
\end{equation}
where $P_{l}^{\text{Tx}}$ is the total radiated power of the \gls{AP} $l$.

\subsubsection{DISTRIBUTED CB PRECODING}
To enable a low-complexity deployment, a distributed precoding strategy based on \gls{CB} is adopted.
In this approach, each \gls{AP} independently computes its precoding vectors using locally acquired channel state information.
This decentralization significantly reduces the computational burden on the \gls{CPU} and minimizes backhaul signaling requirements.
However, as highlighted in~\cite{2017Ngo}, while CB enables practical implementation in \Gls{CF-mMIMO} systems, it inherently sacrifices some performance gains achievable through joint processing and global channel knowledge.
The precoding vector generated by \gls{AP} $l$ for \gls{MS} $k$ is expressed as
\begin{equation}
    \mathbf{w}^{\text{CB}}_{lk} = \sqrt{\rho_{lk}} \frac{\hat{\boldsymbol{h}}_{lk}}{\sqrt{\text{tr}\left(\boldsymbol{R}_{lk} \boldsymbol{\Gamma}_{lk}^{-1} \boldsymbol{R}_{lk}\right)}}.
    \label{eq:cb_precoding_vector}
\end{equation}

For power allocation, each \gls{AP} distributes its available transmit power among the \glspl{MS} employing a fractional policy \cite{2019InterdonatoICC,2020Nikbakht,2021Demir}, defined as
\begin{equation} \label{eq:APA-policy}
{\rho}_{lk}= P_{\mathrm{dl}}\dfrac{\Big(\beta_{lk}\Big)^\nu}
{\sum^K_{\substack{i=1}}\Big(\beta_{li} \Big)^\nu}.
\end{equation}
This allocation ensures that each \gls{AP} transmits at full power as 
\begin{equation}
    P_{l}^{\text{Tx}}=\sum_{k=1}^K \rho_{lk} = P_{\mathrm{dl}}, \quad \forall l \in \mathcal{L}^A.
    \label{eq:ap_power_constraint}
\end{equation}

\subsection{SPATIAL MS DISTRIBUTION MODEL}
\label{sec:Spatial_model}

\begin{figure}
  \centering
  \includegraphics[width=7.8cm]{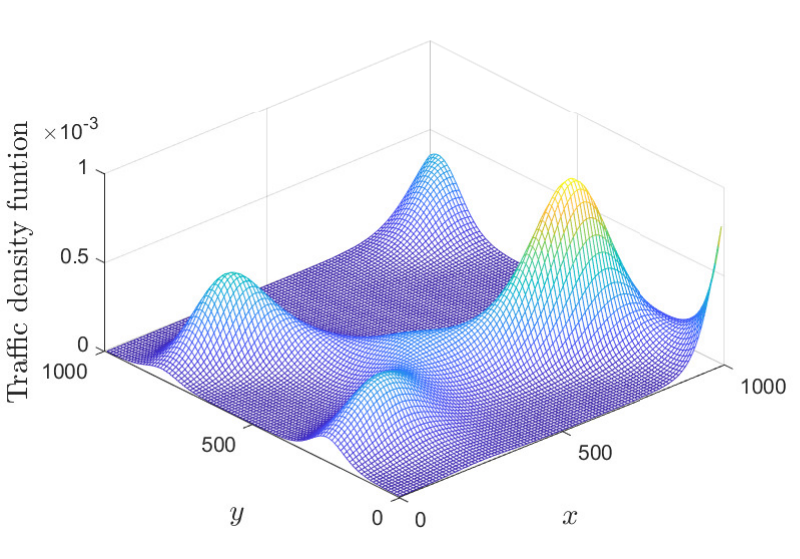}
  \caption{Probability density function of the number of \glspl{MS} per pixel on a square grid of side $1000$ m and square pixels of side $10$ m (see parameters in \cite[Table I]{lee2014spatial} for an urban scenario).}
  \label{fig:MS_pdf}
\end{figure}

In this study, we model the spatial distribution of \glspl{MS} within the service area using the framework proposed by Lee \textit{et al.} in \cite{lee2014spatial}, which was also applied in the context of \gls{CF-mMIMO} in \cite{garcia2020energy}. This spatial traffic model captures large-scale variations in traffic density by using sums of sinusoids, which represent the characteristics of spatially correlated log-normally distributed traffic patterns.

Specifically, a square area denoted by $\mathcal{S}$ is divided into a regular grid of $N_X$ by $N_Y$ rectangular pixels. Each pixel, indexed by $(x, y)$ with $x \in \{1, \dots, N_X\}$ and $y \in \{1, \dots, N_Y\}$, is associated with a traffic density $\rho_{x,y}$ (measured in \glspl{MS} per pixel). For an urban scenario, we generate the traffic density map $\rho_{x,y}$ using a combination of lognormal distributions, as illustrated in Fig.~\ref{fig:MS_pdf}, following the approach detailed in~\cite{lee2014spatial}.

The \gls{PDF} of the number of \glspl{MS} per pixel, denoted by $f^{\gls{MS}}_{x,y}$, can be computed as
\begin{equation} f^{\gls{MS}}_{x,y} = \frac{\rho_{x,y}}{\sum_{x'=1}^{N_X} \sum_{y'=1}^{N_Y} \rho_{x',y'}}. \label{eq:MS_pdf} \end{equation}
This formulation effectively normalizes the traffic distribution across the service area.

\section{PERFORMANCE METRICS}
\label{sec:Performance_metrics}
Let us consider energy efficiency as the key metric for selecting active \glspl{AP} in $\mathcal{L}^A$.
In this section, we first define the sum spectral efficiency and the power consumption model, which will ultimately be used to derive the energy efficiency metric.

\subsection{SPECTRAL EFFICIENCY}
\label{sec:Achievable_rates}

The sum downlink spectral efficiency (measured in bps/Hz) can be expressed as
\begin{equation}
   \Xi(\mathcal{L}^A) =\sum_{k=1}^K {\xi_k(\mathcal{L}^A)},
   \label{eq:SE_fitness}
\end{equation}
where $\xi_k(\mathcal{L}^A)$ is the achievable downlink spectral efficiency of an arbitrary MS $k$, given by \cite{2021Demir}
\begin{equation}
    \xi_k(\mathcal{L}^A)= (\tau_d/\tau_c) \log_2(1 + \gamma_k(\mathcal{L}^A)),
    \label{eq:SE_gk}
\end{equation}
with the effective \gls{SINR} expressed as
\begin{equation}
        \!\gamma_k(\mathcal{L}^A) \!=\! \frac{\left|\sum\limits_{l\in\mathcal{L}^A} \mathbb{E}\{\varrho^{k}_{lk}\} \right|^2}{\sum\limits^K_{i=1} \mathbb{E}\left\{\left|\sum\limits_{l\in\mathcal{L}^A} \varrho^{i}_{lk} \right|^2\right\} \!-\! \left|\sum\limits_{l\in\mathcal{L}^A}{\mathbb{E}\left\{\varrho^{k}_{lk}\right\}}\right|^2 \!\!\!+\! \sigma_d^2},
    \label{eq:SINR_k}
\end{equation}
where $\varrho^{i}_{lk}=\boldsymbol{h}\herm_{lk}  \mathbf{w}^{\text{proc}}_{li}$ denotes the effective downlink channel contribution from \gls{AP} $l$ to \gls{MS} $k$ associated with the data symbol intended for \gls{MS} $i$, with $\mathbf{w}^{\text{proc}}_{li} \in \{\mathbf{w}^{\text{CB}}_{li},\,\mathbf{w}^{\text{MMSE}}_{li}\}$ denoting the processing vector used at \gls{AP}~$l$ for \gls{MS}~$i$.  
The expectations in \eqref{eq:SINR_k} are taken with respect to the random channel realizations.

\subsection{POWER CONSUMPTION MODEL}
\label{sec:power_consumption}
A linear model is used to approximate the total power $P_l^{\text{AP}}$ consumed at the $l$th \gls{AP}, as proposed in several studies \cite{Auer11,Tombaz11,Desset12,Bjornson16c,Dai16}.

In active mode, during the downlink payload data transmission phase, the power consumption at the $l$th \gls{AP} depends on several factors, including the radiated power, the efficiency of the power amplifier, the power consumed by the small-signal \gls{RF} transceiver, baseband circuitry, and losses from components such as the feeder, DC power supply, main supply, and cooling system.

In sleep mode, the \gls{AP} remains in a reduced power state, enabling a relatively quick reactivation. In this mode, although the \gls{AP} is not transmitting or receiving power, certain components such as the power supply, some signal processing blocks, and part of the cooling system remain active, consuming power. The power consumption in each mode is therefore modeled as follows
\begin{equation}
   P_l^{\text{AP}}=\begin{cases}
                      \frac{P_{l}^{\text{Tx}}}{\alpha_l^{\text{AP}}} + P_{l}^{\text{AP,fix}} + N_\text{chain} P_{l}^{\text{AP,chain}} & \text{Active}\\
                      \text{ } & \text{ }\\
                      P_{l,\,\text{sleep}}^{\text{AP,fix}} + N_\text{chain} P_{l,\,\text{sleep}}^{\text{AP,chain}} & \text{Sleep},
                   \end{cases}
\end{equation}
where $\alpha_l^{\text{AP}}$ is the power amplifier efficiency at the $l$th \gls{AP}, $P_{l}^{\text{AP,fix}}$ denotes the downlink power consumption that is independent of both the number of RF chains $N_\text{chain}$ and the traffic load, $P_{l}^{\text{AP,chain}}$ models the downlink traffic-independent power consumed by the circuitry related to each RF chain at the $l$th \gls{AP}, $P_{l,\,\text{sleep}}^{\text{AP,fix}}$ and $P_{l,\,\text{sleep}}^{\text{AP,chain}}$ represent the RF chain-independent and RF chain-dependent power consumed by the $l$th \gls{AP} in sleep mode.

A similar power consumption model can be established for the fronthaul links that connect \glspl{AP} to \gls{CPU}. In particular, the power consumed by the $l$th fronthaul link in active mode depends on the traffic it conveys. The total power consumption is approximated as \cite{Nguyen17,Ngo18}
\begin{equation}
   P_l^{\text{FH}}(\mathcal{L}^A)=\begin{cases}
                      B \, \eta_l^{\text{FH}} \, \Xi(\mathcal{L}^A) + P_l^{\text{FH,fix}} & \text{Active} \\
                      \text{ } & \text{ }\\
                      P_{l,\,\text{sleep}}^{\text{FH,fix}} & \text{Sleep},
                   \end{cases}
\end{equation}
where $B$ is the system bandwidth, $\eta_l^{\text{FH}}$ denotes the traffic-dependent power consumption coefficient (in Watt per bit/s), $P_l^{\text{FH,fix}}$ expresses the traffic-independent power consumption in active mode, and $P_{l,\,\text{sleep}}^{\text{FH,fix}}$ accounts for the power consumed by the $l$th fronthaul link in sleep mode.

For \glspl{MS}, the power consumption during the downlink payload data transmission phase is also approximated by
\begin{equation}
   P_k^{\text{MS}}(\mathcal{L}^A)=B \, \eta_k^{\text{MS}} \xi_k(\mathcal{L}^A) + P_{k}^{\text{MS,fix}} \text{ },
\end{equation}
where $\eta_k^{\text{MS}}$ is the traffic-dependent power consumption coefficient (in Watt per bit/s) for the $k$-th MS and $P_{k}^{\text{MS,fix}}$ models the power consumed by the internal circuitry of the $k$-th \gls{MS}, independent of the radiated power.

By combining the contributions from the \glspl{AP}, fronthaul links, and \glspl{MS}, the total power consumption of the \gls{CF-mMIMO} network during the downlink payload data transmission phase is modeled as
\begin{equation}
\begin{split}
   {P_T(\mathcal{L}^A)} &= P_{T}^{\text{fix}}+B\sum_{k=1}^K \eta_k^{\text{MS}} \xi_k(\mathcal{L}^A)\\ 
   &+\sum_{l=1}^{L_A} \left(\frac{\tau_d}{\tau_c}\frac{P_{l}^{\text{Tx}}}{\alpha_l^{\text{AP}}} + B \, \eta_l^{\text{FH}} \, \Xi(\mathcal{L}^A)\right),
\end{split} 
\end{equation}
 with
\begin{equation}
\begin{split}
    P_{T}^{\text{fix}}&=\frac{\tau_d}{\tau_c}\left[\sum_{k=1}^K P_{k}^{\text{MS,fix}}\right. \\
    &+\sum_{l \in \mathcal{L}^A} \left(P_l^{\text{FH,fix}} + P_{l}^{\text{AP,fix}} + N_\text{chain} P_{l}^{\text{AP,chain}}\right) \\
    &+\left.\sum_{l \in \mathcal{L}^S} \left(P_{l,\,\text{sleep}}^{\text{FH,fix}} + P_{l,\,\text{sleep}}^{\text{AP,fix}} + N_\text{chain} P_{l,\,\text{sleep}}^{\text{AP,chain}}\right)\right].
\end{split} 
\end{equation}

\subsection{ENERGY EFFICIENCY}
\label{sec:energy_efficiency}
The energy efficiency during the downlink payload data transmission phase is defined as the ratio of the system's total spectral efficiency to the total power consumption. This can be mathematically expressed as
\begin{equation}
   \Upsilon(\mathcal{L}^A)=\frac{B \, \Xi(\mathcal{L}^A)}{{P_T(\mathcal{L}^A)}}.
   \label{eq:EE_fitness}
\end{equation}
This metric measures the efficiency with which the system converts transmitted data into a useful payload, relative to its energy consumption. A higher energy efficiency implies that the system is able to deliver more data for less power, which is particularly important in the context of green communications and energy-conscious network design.

\section{EVOLUTIONARY AP SWITCHING STRATEGIES}
\label{sec:GA_ASW}
To optimize energy efficiency, the set of active \glspl{AP} should dynamically adjust to variations in the network scenario. These variations may stem from factors such as changes in the number and/or location of \glspl{MS}, or shifts in the geographical distribution of shadow fading.
To address this challenge, this paper proposes two evolutionary approaches for \gls{ASW}. The first approach utilizes a \gls{CGA}, which delivers a significant improvement in energy efficiency compared to benchmark methods that rely on greedy-search strategies (see, for example, \cite{garcia2020energy} and \cite{femenias2020access}). The second approach, an unconstrained \gls{PDGA}, offers a more comprehensive view by presenting energy-spectral efficiency trade-offs as a Pareto front.

The optimization problem can be formalized as 
\begin{equation}
\label{eq:Optimal} 
{\mathcal{L}^A}^* = \arg\max_{\substack{\mathcal{L}^A \subseteq \mathcal{L}}} \Upsilon(\mathcal{L}^A ),
\end{equation} 
where ${\mathcal{L}^A}^*$ represents the optimal subset of active \glspl{AP} that maximizes the overall energy efficiency.
Solving the optimization problem in equation \eqref{eq:Optimal} is NP-hard, as it requires evaluating the performance of all possible combinations of $L_A$ out of $L$ \glspl{AP}. Consequently, computationally efficient search strategies, such as those based on heuristics or evolutionary algorithms, are necessary for practical implementation, which is the focus of the following discussion.

\subsection{CONSTRAINED GENETIC ALGORITHM (CGA)}
\label{sec:CGA_ASW}
To determine the subset of active \glspl{AP} ($\mathcal{L}^A$) that maximizes energy efficiency, depending on the number of active \glspl{AP} ($L_A$), we propose a \gls{CGA} approach.
The decision to use a \gls{GA} \cite{holland1975adaptation} stems from the nature of the problem, which is well-suited for such optimization techniques. Specifically, the challenge involves maximizing a non-deterministic fitness function, where the number of possible combinations of active \glspl{AP} is prohibitively large for exhaustive search methods. For example, if $L_A = 6$ are selected from $L = 100$ available \glspl{AP}, the number of combinations is over a billion. Thus, a GA is a natural choice for efficiently navigating the solution space.

When using \gls{CGA}, the core structure of the \gls{GA} is modified to impose a constraint on the number of active \glspl{AP} ($L_A$) at different stages of the algorithm. Specifically, for a value of $L_A$, the \gls{GA} seeks the combination of active \glspl{AP} that maximizes the energy efficiency function $\Upsilon(\mathcal{L}^A )$.
This constraint is applied during the initial population generation, crossover, and mutation operations, ensuring that all individuals in the population correspond to solutions with exactly $L_A$ active \glspl{AP}.
The details of the \gls{CGA} implementation are outlined in Algorithm \ref{alg:CGA}.
The proposed algorithm focuses on identifying the most \gls{EE} subset of \glspl{AP} under a fixed cardinality constraint.
The procedure begins by generating an initial population of candidate solutions, where each individual activates exactly $L_A$ access points.
At each generation, the \gls{EE} of all individuals is evaluated using the fitness function, and an elitist selection mechanism retains the best-performing fraction of the population.
New candidate solutions are then produced through crossover among the selected individuals, while explicitly enforcing the constraint on the number of active \glspl{AP}. Additionally, mutation is applied with a prescribed probability to introduce small random perturbations and enhance the exploration of the solution space.
The next generation is formed by combining the elitist survivors, the offspring generated by crossover, and randomly created individuals, all of which satisfy the $L_A$-\gls{AP} constraint. This iterative process continues for a predefined number of generations, ultimately yielding a near-optimal or suboptimal set of active \glspl{AP} $\mathcal{L}^A$ that maximizes the \gls{EE} for the given value of $L_A$.

\begin{algorithm}
\caption{Proposed constrained genetic algorithm (CGA)}
\label{alg:CGA}

\begin{algorithmic}

\STATE \textbf{Inputs:}
\STATE - \Gls{CF-mMIMO} system parameters (e.g., $L$, $K$),
\STATE - Randomly deploy $L$ \glspl{AP}.
\STATE - Spatial \gls{MS} distribution (in terms of $f^{\gls{MS}}_{x,y}$, $\forall x,y$).

\STATE \textbf{Restrictions:}
\STATE - Number of active \glspl{AP} ($L_A$).

\STATE \textbf{Initializations:}
\STATE - Generate the initial population of \( N_{\text{eval}} \) individuals, ensuring that each individual has exactly $L_A$ active \glspl{AP} (randomly selected).

\WHILE{\textit{number of generations} \( \leq G \)}
\STATE - Evaluate the fitness of each individual using the \gls{CF-mMIMO} fitness function \eqref{eq:EE_fitness} to obtain the the energy efficiency value for each individual.
\STATE - Apply elitist selection, retaining the best $p_e$ of the individuals based on the energy efficiency value (i.e., select the top $p_e N_{\text{eval}}$ chromosomes). 
\STATE - Perform crossover with a probability $p_c$, generating $p_c p_e N_{\text{eval}}$ offspring by combining selected individuals.
\STATE - Ensure that the new offspring satisfy the constraint of having exactly $L_A$ active \glspl{AP}.
\STATE - Apply mutation with a probability $p_m$, introducing small random changes in the individual to explore the solution space.
\STATE - Generate a new population by merging the selected individuals (the best one of elitist selection and individuals generated by crossover) and the newly generated offspring (randomly generated), ensuring that each individual has exactly $L_A$ active \glspl{AP}.
\ENDWHILE

\STATE \textbf{Outputs:}
\STATE - Optimal/near-optimal $\mathcal{L}^A$ set that maximizes $\Upsilon$ while being restricted to $|\mathcal{L}^A|=L_A$.

\end{algorithmic}
\end{algorithm}


\vspace{0.2cm}
{\bf CGA computational complexity analysis:}

The computational complexity of the CGA depends on several factors \cite{papadimitriou2003computational,cook2007overview}: the number of generations $G$, the number of the newly evaluated individuals (chromosomes) per iteration \(N_{\text{eval}}\), the number of elements in each individual (which corresponds to the number of active \glspl{AP} $L_A$), and the evaluation cost of the fitness function, which varies based on $L_A$ and the system parameters of the \gls{CF-mMIMO} network when calculating energy efficiency.

Let $f(L_A)$ represent the fitness function evaluation cost as a function of the number of active \glspl{AP} $L_A$, while keeping the other CF-mMIMO system parameters constant.
After the initialization step, the computational complexity for each generation is broken down as follows:
\begin{itemize}
\item Evaluation of individuals: the evaluation of $N_{\text{eval}}$ individuals, each requiring the calculation of the fitness function, has a complexity of \(\mathcal{O}(N_{\text{eval}} \ f(L_A))\). 

\item Selection of the best individuals: the selection process involves sorting the \( N_{\text{eval}} \) individuals based on their fitness values. Sorting a vector in descending order to rank the individuals in terms of energy efficiency has a complexity of \(\mathcal{O}(N_{\text{eval}} \log_2({N_{\text{eval}}}))\).

\item Constrained crossover: this operation involves recombining the best individuals to create offspring. In this case, \( L_A \) elements are transferred from parents to offspring. Given that the number of active \glspl{AP} \( L_A \) can be scaled as a function of \( L \), the crossover step has a complexity of \(\mathcal{O}(L \ N_{\text{eval}})\).

\item Constrained mutation: similar to the crossover operation, mutation is applied to individuals, introducing small random changes while respecting the constraint on \( L_A \). This operation also has a complexity of \(O(L \ N_{\text{eval}})\).
\end{itemize}
Thus, the computational complexity across $G$ generations is given by 
\begin{equation}
\mathcal{O}\left(G \ N_{\text{eval}} \ \max\left\{f(L_A), \ L, \ \log N_{\text{eval}}\right\}\right).
\end{equation}


Algorithm~\ref{alg:CGA} determines the subset of active \glspl{AP} that maximizes \gls{EE} for a given value of $L_A$.
By running the algorithm for multiple values of $L_A$, with $1 \leq L_A \leq L$, and \gls{EE} curve as a function of $L_A$ can be obtained, and its maximizing point can be identified.
For a single execution, the computational cost of the fitness function, denoted as $f(L_A)$, depends explicitly on the chosen value of $L_A$.
However, when the full \gls{EE} curve is computed, the total number of fitness evaluations scales proportionally with $L$.
Therefore, for asymptotic analysis, it is appropriate to express the fitness function cost in terms of $L$.
Accordingly, the overall computational complexity can be characterized as
\begin{equation}
\mathcal{O}\left(L \ G \ N_{\text{eval}} \ \max\left\{f(L), \ L, \ \log N_{\text{eval}}\right\}\right).
\end{equation}

\vspace{0.2cm}
{\bf Analysis of dominance (CGA):}

\begin{enumerate}
    \item \textbf{Fitness evaluation} (\( \mathcal{O}(L \ G \ N_{\text{eval}} \ f(L)) \)): 
    This term dominates when the cost \(f(L)\) is computationally expensive. In practical scenarios, such as in \textit{near-optimal AP green-mode solution} for \gls{CF-mMIMO}, evaluating the energy efficiency requires detailed system modeling and simulations, making \(f(L)\) the most computationally intensive component.
    However, the fixed number of evaluations \(N_{\text{eval}}\) and the elitist selection strategy mitigate the impact of this term, allowing the computational complexity of the algorithm to remain affordable even with high-cost fitness evaluations, since independent optimization runs can be performed in parallel. 
    
    \item \textbf{Crossover and mutation} (\( \mathcal{O}(G \ L^2 \ N_{\text{eval}}) \)):  
    As this term  scales quadratically with \(L\), these operations are more significant when the solution representation length \(L\) is large, such as in scenarios involving many available \glspl{AP}. However, since the computational cost of crossover and mutation is generally lower than the cost of fitness evaluations, their contribution to the overall complexity is typically negligible. Therefore, these steps are rarely the bottleneck in most use cases.
    
    \item \textbf{Best individuals selection} (\( \mathcal{O}(L \ G \ N_{\text{eval}} \ \log_2({N_{\text{eval}}})) \)):  
    This term accounts for the computational cost of repeatedly sorting a vector of length \( N_{\text{eval}} \) in descending order, across \(G\) generations and $L_\mathcal{V}$ evaluations. However, given that genetic algorithms typically operate with relatively small values of \( N_{\text{eval}} \), the impact of this term on the overall complexity is also considered negligible.
\end{enumerate}

\subsection{PARETO-DRIVEN GENETIC ALGORITHM (PDGA)}
\label{sec:PDGA_ASW}
The \gls{PDGA} is a multi-objective optimization technique designed to balance spectral efficiency \(\Xi\) and energy efficiency \(\Upsilon\) in network configurations. It is optimized for computational efficiency by operating with a relatively small population and evaluating a fixed number of new solutions per iteration. Its core mechanism identifies and preserves non-dominated solutions, guiding the search toward optimal trade-offs between competing objectives.

\begin{algorithm}
\caption{Pareto-driven genetic algorithm (\gls{PDGA})}
\label{alg:PDGA}

\begin{algorithmic}

\STATE \textbf{Inputs:}
\STATE - \Gls{CF-mMIMO} system parameters (e.g., $L$, $K$),
\STATE - Randomly deploy $L$ \glspl{AP}.
\STATE - Spatial \gls{MS} distribution (in terms of $f^{\gls{MS}}_{x,y}$, $\forall x,y$).

\STATE \textbf{Restrictions:} No specific restrictions.

\STATE \textbf{Initializations:}
\STATE - Generate the initial population of $N_{\text{eval}}$ individuals, ensuring inclusion of an 'all-APs-ON' individual known for achieving the highest spectral efficiency.
\STATE - Evaluate the fitness of the initial population using the \gls{CF-mMIMO} fitness functions \eqref{eq:SE_fitness} and \eqref{eq:EE_fitness} to obtain $\Xi$ and $\Upsilon$ values, respectively.

\WHILE{\textit{number of generations} \( \leq G \)}
\STATE - Identify individuals that belong to the Pareto front and discard the rest (selection).

\STATE - Generate new offspring through crossover ($N_{\text{eval}}$ individuals) as:
\begin{itemize}
\STATE \hspace{1em}50\% chance: select the first parent from the top $N_{top}$ individuals with the highest $\Upsilon$ or select it randomly from the surviving population.
\STATE \hspace{1em}50\% chance: select the second parent from the top $N_{\text{top}}$ individuals with the highest $\Upsilon$ or select it randomly from the surviving population.
\end{itemize}

\STATE - Apply mutation to the offspring with the following probabilities:
\begin{itemize}
\STATE \hspace{1em} With probability \(p_{m1}\): set one randomly selected element to $\text{OFF}$.
\STATE \hspace{1em} With probability \(p_{m2}\): set a fraction $r_{m2}$ of the elements to $\text{OFF}$.
\STATE \hspace{1em} With probability \(p_{m3}\): perform $N_{\text{swp}}$ random swaps within the individual.
\end{itemize}

\STATE - Evaluate the fitness of the new population to calculate $\Xi$ and $\Upsilon$ values.

\STATE - Merge the surviving individuals and the newly generated individuals to form the next generation's population.
\ENDWHILE

\STATE \textbf{Outputs:}
\STATE - Optimal/suboptimal $\mathcal{L}^A$ set corresponding to the best $\Xi$ and $\Upsilon$ values (Pareto front representing the trade-off between spectral and energy efficiencies).

\end{algorithmic}
\end{algorithm}

Building on seminal works in Pareto-based optimization~\cite{Srinivas1994,Deb2002,Horn1994,Faghihi2016}, the \gls{PDGA} adopts an elitist selection strategy that retains the best individuals on the Pareto front. This accelerates convergence toward the optimal Pareto set while allocating evaluations more effectively. Insights from niched Pareto methods~\cite{Horn1994} and multi-objective scheduling applications~\cite{Faghihi2016} further support its applicability to complex optimization tasks.

To preserve solution diversity, \gls{PDGA} employs a \emph{uniform crossover} operator (50/50, gene-wise) together with probabilistic \emph{mutation} strategies. The latter include bit flipping, zeroing out a fraction of elements, and performing random swaps (exchanging the activation state of two randomly selected \glspl{AP})~\cite{Shimodaira1996,Arabas1998,Liu2016}. In our design, these mutations are intentionally \emph{biased toward deactivating} \glspl{AP} (i.e., toward sparser activation patterns), which accelerates the discovery of energy-efficient configurations. Consistently with this goal, parent selection is \emph{biased toward high \(\Upsilon\)} (e.g., selecting from the top \(N_{\text{top}}\) individuals with higher energy efficiency with a given probability), thereby reinforcing the search pressure toward energy savings.

At the same time, to avoid a unidirectional drift that would excessively remove active \glspl{AP} and harm spectral efficiency, we \emph{inject and retain} the 'all-APs-ON' individual in the population. This chromosome acts as a persistent source of ones in a system whose mutations tend to eliminate them, ensuring coverage of the high \(\Xi\) region and preserving Pareto diversity. Taken together, these mechanisms enable a comprehensive exploration of the solution space while maintaining a careful balance between spectral and energy efficiency trade-offs. The detailed implementation of the \gls{PDGA} is presented in Algorithm~\ref{alg:PDGA}.
The proposed algorithm starts by generating an initial population of candidate solutions, including a reference individual where all access points are active to ensure a maximum spectral efficiency baseline. Then, at each generation, individuals belonging to the Pareto front are selected, while the rest are discarded. New individuals are generated through crossover between selected parents, with a bias towards high-energy-efficiency solutions, and mutation is applied to introduce diversity by randomly switching off access points or swapping their states. The fitness of the new population is evaluated in terms of spectral efficiency and energy efficiency, and both the surviving and newly generated individuals are merged to form the next generation. This process is repeated for a predefined number of generations, and the final Pareto front provides the set of optimal and suboptimal solutions representing the trade-off between spectral efficiency and energy efficiency.

It is important to remark that the objective of the proposed \gls{PDGA}-based approach is not limited to generating a set of candidate \gls{AP} activation patterns, but rather to explicitly characterize the trade-off between energy and spectral efficiencies in the considered \gls{CF-mMIMO} system.
The multi-objective formulation identifies the set of non-dominated solutions spanning the achievable \gls{EE}–\gls{SE} region, providing insight into inherent trade-offs, including regimes where \gls{SE} gains require disproportionately higher energy consumption and others where small \gls{SE} reductions yield significant energy savings.
This enables a decision-oriented interpretation of the \gls{AP} ON/OFF problem, allowing the selection of operating points based on system-level requirements, such as minimum \gls{SE} or QoS constraints, thereby transforming it into a flexible decision-support tool for \gls{CF-mMIMO} network configuration.
This aspect will be further illustrated in Section~\ref{sec:numerical_results}, where the Pareto front is exploited to select representative operating points.

\vspace{0.2cm}
{\bf \gls{PDGA} computational complexity analysis:}

The computational complexity of \gls{PDGA} is influenced by several factors: the number of generations $G$, the fixed number of newly evaluated individuals per iteration $N_{\text{eval}}$, the length of the solution representation (the total number of available \glspl{AP} $L$), the cost of fitness function evaluations, and the number of objectives $N_{\text{obj}}$ that affect the identification of the Pareto front \cite{papadimitriou2003computational,cook2007overview}. Given that the total number of individuals is constrained to $G N_{\text{eval}}$, the primary focus of the analysis is on $G$. Additionally, the fitness evaluation cost depends on the number of active \glspl{AP} $L_A$ (while the remaining system parameters of the \gls{CF-mMIMO} are held constant).
As previously stated, the possible values of $L_A$ scale with $L$, and aiming to determinate computational complexity order, the cost of the fitness function is characterized as a funtion of $L$.
 
The computational complexity of the \gls{PDGA} can be analyzed by examining the main operations performed at each generation. 
At every iteration, a fixed number of new candidate solutions ($N_{\text{eval}}$) are generated and evaluated, while the current set of non-dominated solutions is updated.

\begin{itemize}
    \item \textbf{Fitness evaluation:} the evaluation of $N_{\text{eval}}$ individuals requires computing the \gls{CF-mMIMO} fitness functions, leading to a complexity of \(\mathcal{O}(N_{\text{eval}} \ f(L))\).

    \item \textbf{Pareto front identification:} the identification of non-dominated solutions requires pairwise comparisons among candidate solutions across generations. 
    In the worst case, the number of individuals involved in the dominance checks may grow proportionally to the number of generations, which leads to a per-generation complexity of 
    \(\mathcal{O}(N_{\text{eval}}^2 \ G^2 \ N_{\text{obj}})\).

    \item \textbf{Crossover:} generating offspring through crossover requires transferring elements of the chromosome of length $L$, leading to a complexity of \(\mathcal{O}(L \ N_{\text{eval}})\).

    \item \textbf{Mutation:} mutation operations modify elements of the solution representation, resulting in a complexity of \(\mathcal{O}(L \ N_{\text{eval}})\).
\end{itemize}

Thus, the overall computational complexity across $G$ generations can be expressed as

\begin{equation}
\mathcal{O}\left(G \ N_{\text{eval}} \ \max\left\{f(L), \ L, \ N_{\text{eval}} G^2 N_{\text{obj}} \right\}\right).
\end{equation}

\vspace{0.2cm}
{\bf Analysis of dominance (\gls{PDGA}):}

\begin{enumerate}
    \item \textbf{Fitness evaluation} (\( \mathcal{O}(G N_{\text{eval}} \ f(L)) \)):  
    This term accounts for the repeated evaluation of the fitness function across generations and individuals. 
    In practical \gls{CF-mMIMO} scenarios, this step typically dominates the computational cost since evaluating the energy and spectral efficiency requires detailed system-level simulations.

    \item \textbf{Crossover and mutation} (\( \mathcal{O}(G \ L \ N_{\text{eval}}) \)):  
    These operations scale linearly with the length of the solution representation ($L$). 
    Compared to CGA, where the chromosome structure is constrained by the number of active \glspl{AP}, the PDGA representation operates directly on the full set of available \glspl{AP}, leading to a simpler linear scaling.

    \item \textbf{Pareto front identification} (\( \mathcal{O}(N_{\text{eval}}^2 G^3 N_{\text{obj}}) \)):  
    This term reflects the cumulative cost of identifying non-dominated solutions across generations. 
    As the number of candidate solutions may grow with the number of generations, the worst-case complexity scales quadratically with \(N_{\text{eval}}\), cubically with \(G\), and linearly with the number of objectives \(N_{\text{obj}}\). 
    Although this term can become significant in the worst case, the dominance checks involve simple comparisons and are generally less computationally demanding than the fitness evaluations required in \gls{CF-mMIMO} simulations.
\end{enumerate}

\section{BENCHMARK GREEDY SEARCH-BASED \gls{ASW}}
\label{sec:Gready_ASW}
An exhaustive greedy search-based \gls{ASW} strategy (referred to as benchmark) was proposed in \cite{garcia2020energy} to maximize average energy efficiency.
This benchmark strategy implements an iterative algorithm that incorporates \gls{GoF} assistance by using different methods, such as, Chi-square, Kolmogorov-Smirnov, and \textit{statistical energy} \gls{GoF} tests.
The \gls{GoF} assistance has been specifically designed to cope with long-term non-uniform spatial traffic densities.
The study in \cite{garcia2020energy} demonstrates the benefits of \gls{GoF} methods when trying to match the spatial distribution of active \glspl{AP} to that of \glspl{MS} in the network. This approach aims to selectively activate parts of the network where active \glspl{MS} are most likely located.
More specifically, when a particular probability distribution has been specified to model a random phenomenon (such as the spatial distribution of \glspl{MS}), the validity of the assumed distribution model can be statistically verified or disproven by using \gls{GoF} tests \cite{dAgostino86,aslan2002comparison,evans2008distribution,Williams10,read2012goodness}.
In this context, for a given number of \glspl{AP} to be active ($L_A$), \gls{GoF} techniques can determine which \glspl{AP} should be turned on or off in such a way that the resulting active \gls{AP} distribution matches the non-uniform \gls{MS} distribution.

As previously described in Section \ref{sec:Spatial_model}, we assume that the target region is tessellated into a regular grid of $N_X$ by $N_Y$ rectangular pixels, and the probability density of \glspl{MS} in pixel $(x,y)$ is denoted as $f_{x,y}^{MS}$. For a given set $\mathcal{L}^A$ of active \glspl{AP}, we can estimate the probability density of \glspl{AP}, $f_{x,y}^{AP}$, for this particular pixel as
\begin{equation}
   f_{x,y}^{\gls{AP}}=\frac{L_A^{(x,y)}}{L_A},
   \label{eq:APpdf}
\end{equation}
where $L_A^{(x,y)}$ is the number of active \glspl{AP} in pixel $(x,y)$, and $L_A$ is the number of active \glspl{AP} in the target region. This relationship establishes a link between the spatial distribution of \glspl{MS} ($f_{x,y}^{\gls{MS}}$) and active \glspl{AP} ($f_{x,y}^{\gls{AP}}$) and is essential for GoF techniques. 
These techniques help optimize the placement of \glspl{AP} by evaluating a \emph{discrepancy (or misfit) metric} to achieve the best alignment between the distributions of active \glspl{AP} and \glspl{MS}.
The discrepancy metric  $D_{\mathcal{L}^A}$ can be expressed as a function of $f_{x,y}^{\gls{MS}}$ and $f_{x,y}^{\gls{AP}}$ and calculated as in \cite{garcia2020energy} for different \gls{GoF} tests.

Using these definitions, a greedy algorithm is implemented. Starting with a set that contains all the \glspl{AP} in the ON mode, the algorithm iteratively switches off a single \gls{AP} resulting in the minimum discrepancy metric value when that specific \gls{AP} is deactivated.
In this way, for a given number of active \glspl{AP} (in ON mode), the algorithm identifies which ones should be turned off.

To explain the benchmark search, let us first define a system state $s$ as a given number of active \glspl{AP} ($s = L_A$).
For example, the system is in the state $L$ ($s = L$) when all available \glspl{AP} in the network are active.
Let us also define $\mathcal{L}^A_{s}$ as the set $\mathcal{L}^A$ for $s$ \glspl{AP} in ON mode.
This strategy addresses the optimization problem defined in equation~\eqref{eq:Optimal} by first determining $\mathcal{L}^A_{s}$ for each state $s$, and subsequently solving the next optimization problem in the final steps as
\begin{equation}
\label{eq:OptimalM}
s^* = \arg\max_{\substack{s}} \Upsilon(s,\mathcal{L}^A_{s}|L,f^{\gls{MS}}_{x,y}),
\end{equation}
where $s^*$ is the optimal system state (i.e., the optimal number of active \glspl{AP}), and $\Upsilon(s,\mathcal{L}^A_{s}|L,f^{\gls{MS}}_{x,y})$ represents the energy efficiency as a function of the system state $s$ and the set $\mathcal{L}^A_{s}$ of \glspl{AP} in ON mode for state $s$, under a given scenario with $L$ available \glspl{AP} and following a given \gls{MS} distribution (in terms of $f^{\gls{MS}}_{x,y}$, $\forall x,y$).
Notice that the number $K$ of \glspl{MS} and the rest of the \gls{CF-mMIMO} system parameters are assumed to be constant.
From this point forward, for simplicity, we will refer to energy efficiency for a given scenario as a function of the system state $s$ and the set $\mathcal{L}^A_{s}$, simply as $\Upsilon(s,\mathcal{L}^A_{s})$.
A mathematical pseudocode for the benchmark iterative algorithm is shown in Algorithm \ref{alg:Gready}.

\begin{algorithm}
\caption{Benchmark greedy search algorithm}
\label{alg:Gready}

\begin{algorithmic}

\STATE \textbf{Inputs:}
\STATE - \Gls{CF-mMIMO} system parameters (e.g., $L$, $K$, $\beta_{lk} \forall l,k$),
\STATE - Spatial \gls{MS} distribution (in terms of $f^{\gls{MS}}_{x,y}$, $\forall x,y$).

\STATE \textbf{Initializations:}
\STATE - Deploy $L$ \glspl{AP} randomly.
\STATE - Set $s = L$ and $\mathcal{L}^A_{s} = \mathcal{L}$ (all available \glspl{AP} in ON mode).
\STATE - Obtain $\Upsilon(s,\mathcal{L}^A_{s})$
\STATE - Save $\Upsilon(s,\mathcal{L}^A_{s})$ and $\mathcal{L}^A_{s}$ for the state $s = L$.

\FOR{$s=(L-1)$ down-to $1$ system states}
\STATE - Call \gls{GoF} assistance subroutine (\textbf{Algorithm \ref{alg:GoF}}) for $s$.
\STATE - Set  $\mathcal{L}^A_{s} = \mathcal{L}^A$

\STATE - Obtain $\Upsilon(s,\mathcal{L}^A_{s})$
\STATE - Save $\Upsilon(s,\mathcal{L}^A_{s})$ and $\mathcal{L}^A_{s}$ for current state $s$.
\ENDFOR

\STATE \textbf{Final steps:}
\STATE - Using $\Upsilon(s,\mathcal{L}^A_{s}), \forall s$, compute $s^*$ from \eqref{eq:OptimalM}.
\STATE - Obtain $\mathcal{L}^A$ for $s^*$ \glspl{AP} in ON mode as $\mathcal{L}^A=\mathcal{L}^A_{s^*}$.

\STATE \textbf{Outputs:}
\STATE - Optimal/suboptimal $\mathcal{L}^A$ set that maximizes energy efficiency.

\end{algorithmic}
\end{algorithm}

\begin{algorithm}
\caption{\gls{GoF} assistance subroutine}
\label{alg:GoF}

\begin{algorithmic}

\STATE \textbf{Inputs:}
\STATE - The locations of \glspl{AP},
\STATE - The spatial \gls{MS} distribution,
\STATE - The $s$ value.

\STATE \textbf{Initializations:}
\STATE - Set \glspl{AP} of $\mathcal{L}^A_{s+1}$ in ON mode.

\STATE \textbf{Do:}
\STATE - Switch off a single \gls{AP} resulting in the minimum discrepancy metric $D_{\mathcal{L}^A}$.

\STATE \textbf{Outputs:}
\STATE - The set $\mathcal{L}^A$ of $s$ \glspl{AP} in ON mode.

\end{algorithmic}
\end{algorithm}

In Algorithm \ref{alg:Gready}, the $L$ \glspl{AP} are randomly deployed at the beginning.
For each system state $s$, Algorithm \ref{alg:Gready} calls the \gls{GoF} assistance subroutine (see Algorithm \ref{alg:GoF}).
By using this subroutine, the set of active \glspl{AP} that best matches the \gls{MS} distribution is obtained for a given system state $s$. 
The \gls{CF-mMIMO} fitness function is embedded in this algorithm, which includes the deployment of \glspl{MS} and the computation of $\Upsilon(s,\mathcal{L}^A_{s})$ for each state $s$.
Once the optimal value $s^*$ is found by using \eqref{eq:OptimalM}, the algorithm determines which \glspl{AP} should remain in ON mode for energy efficiency maximization.

We can note that Algorithm \ref{alg:Gready} employs the greedy-search method in an iterative manner to avoid running into an NP-hard problem, while assistance Algorithm \ref{alg:GoF} uses very large-scale system-state information: the spatial \gls{MS} distribution.

\vspace{0.2cm}
\textbf{Benchmark computational complexity analysis:}

The computational complexity analysis of the benchmark greedy search algorithm is performed based on the number of evaluations and the cost of evaluating the \gls{CF-mMIMO} fitness function. Since the fitness function evaluation cost depends on the number of active \glspl{AP} (i.e., depends on the system state $s$), and the possible values of $s$ scale with $L$, the computational complexity is given by
\begin{equation}
\mathcal{O}\left(L \ \max\left\{f(L), \ C_{D_{\mathcal{L}^A}} \right\}\right),
\end{equation}
where $C_{D_{\mathcal{L}^A}}$ represents the cost of calculating the minimum discrepancy metric during the \gls{GoF} assistance subroutine. This analysis highlights the importance of optimizing both the fitness function evaluation and the discrepancy metric computation, as both factors directly affect the execution time of the algorithm.

\section{NUMERICAL RESULTS}
\label{sec:numerical_results}
A comprehensive set of numerical results are presented in this section to evaluate the impact of the proposed evolutionary \gls{ASW} strategies on both energy and spectral efficiencies.
To this end, Monte Carlo simulations are performed using the canonical \gls{CF-mMIMO} network model described in Section~\ref{sec:System_model}, in which all active single-antenna \glspl{AP} concurrently serve every \gls{MS} in the coverage area.
The pilot allocation is configured as explained in~\cite{2021Demir}.
A realistic spatial distribution of \glspl{MS}, as detailed in Section \ref{sec:Spatial_model}, is assumed following a combination of lognormal distributions, as illustrated in Fig.~\ref{fig:MS_pdf}. 
The main system parameters used in all simulations are summarized in Table~\ref{tab:parameters}, adapted from previous studies (e.g., \cite{Auer11,Bjornson16c,2017Ngo,Nguyen17,3GPP17,Bjornson19}).

\begin{table}[t!]
\renewcommand{\arraystretch}{0.85}
\caption{\small Cell-free massive MIMO simulation parameters}
\label{tab:parameters}
\centering
\begin{tabular}{l|c}
\hline
\bfseries Parameters & \bfseries Value\\
\hline
\footnotesize {Carrier frequency} & \footnotesize {2 GHz}\\
\footnotesize {Bandwidth ($B$)} & \footnotesize {20 MHz}\\
\footnotesize {Side of the square coverage area} & \footnotesize {1000 m}\\
\footnotesize {AP/MS antenna height} & \footnotesize {10/1.65 m}\\
\footnotesize {Number of available \glspl{AP} ($L$)} & \footnotesize {100}\\
\footnotesize {Number of \glspl{MS} in the network ($K$)} & \footnotesize {20}\\
\footnotesize {Noise figure} & \footnotesize {7 dB} \\
\footnotesize {Power spectral density of noise} & \footnotesize {-174 dBm/Hz}\\
\footnotesize {AP maximum transmit power ($P_{\mathrm{dl}}$)} & \footnotesize {200 mW}\\
\footnotesize {MS maximum transmit power ($P_{\mathrm{ul}}=P_p$)} & \footnotesize {100 mW}\\
\footnotesize {Power control coefficients ($\nu, \kappa$)} & \footnotesize {-0.5, 0.5}\\
\footnotesize {Coherence interval length ($\tau_c$)} & \footnotesize {200 samples}\\
\footnotesize {Training phase length ($\tau_p$)} & \footnotesize {20 samples}\\
\footnotesize {Pathloss model} & \footnotesize{30.5+36.7$\log_{10}(d)$} \\
\footnotesize {Shadow fading decorrelation distance} & \footnotesize {9 m}\\
\footnotesize {Shadow standard deviation} & \footnotesize {4 dB}\\
\footnotesize {Power amplifier efficiency ($\alpha_l^{\text{AP}}$}) & \footnotesize {0.39}\\
\footnotesize {Power per \gls{MS} traffic ($\eta_k^{\text{MS}}$}) & \footnotesize {0.25 W/Gbps}\\
\footnotesize {Power per FH traffic ($\eta_l^{\text{FH}}$)} & \footnotesize {0.25 W/Gbps}\\
\footnotesize {\gls{AP} fixed power ($P_{l}^{\text{AP,fix}}$)} & \footnotesize {6 W}\\
\footnotesize {\gls{AP} fixed power per RF chain ($P_{l}^{\text{AP,chain}}$)} & \footnotesize {0.2 W}\\
\footnotesize {\gls{AP} fixed-sleep power ($P_{l\,\text{sleep}}^{\text{AP,fix}}$)} & \footnotesize {0.8 W}\\
\footnotesize {\gls{AP} fixed-sleep power per RF chain ($P_{l\,\text{sleep}}^{\text{AP,chain}}$)} & \footnotesize {0.02 W}\\
\footnotesize {\gls{MS} fixed power ($P_{k}^{\text{MS,fix}}$)} & \footnotesize {0.75 W}\\
\footnotesize {FH fixed power ($P_l^{\text{FH,fix}}$)} & \footnotesize {4 W}\\
\footnotesize {FH fixed-sleep power ($P_{l\,\text{sleep}}^{\text{FH,fix}}$)} & \footnotesize {0.5 W}\\
\footnotesize {Benchmark \gls{GoF} assistance test} & \footnotesize {Chi-square}\\
\hline
\end{tabular}
\end{table}

\begin{table}[t!]
\renewcommand{\arraystretch}{0.85}
\caption{\small CGA-\gls{ASW} algorithm implemented parameters}
\centering
\begin{tabular}{l|c|c}
\hline
\textbf{Variable} & \textbf{Symbol} & \textbf{Value} \\
\hline
Number of generations & $G$ & 100 \\
Evaluated individuals per iteration & $N_{\text{eval}}$ & 50 \\
Elitism rate & $p_e$ & 0.10 \\ 
Crossover probability & $p_c$ & 0.80 \\ 
Mutation probability & $p_m$ & 0.05 \\ 
\hline
\end{tabular}
\label{tab:CGA-params}
\end{table}

\begin{table}[t!]
\renewcommand{\arraystretch}{0.85}
\caption{\small PDGA-\gls{ASW} algorithm implemented parameters}
\label{tab:PDGA-params}
\centering
\begin{tabular}{l|c|c}
\hline
\textbf{Variable} & \textbf{Symbol} & \textbf{Value} \\
\hline
Number of generations & $G$ & 500 \\
Evaluated individuals per iteration & $N_{\text{eval}}$ & 10 \\
Number of objectives & $N_{\text{obj}}$ & 2 \\ 
Top individuals prioritized in crossover & $N_{\text{top}}$ & 6 \\ 
Mutation prob.: one AP set to Off & $p_{m1}$ & 0.30 \\ 
Mutation prob.: fraction of \glspl{AP} set to Off & $p_{m2}$ & 0.35 \\ 
Fraction of \glspl{AP} set to Off (for $p_{m2}$) & $r_{m2}$ & 0.10 \\ 
Mutation prob.: random swaps & $p_{m3}$ & 0.35 \\ 
Number of swaps (in swap mutation) & $N_{\text{swp}}$ & 5 \\ 
\hline
\end{tabular}
\end{table}

In this study, for the proposed CGA Algorithm \ref{alg:CGA}, we have selected a population size of $N_{\text{eval}}=50$ chromosomes, a maximum number of generations $G=100$, an elitist selection of the best $p_e=0.1$ of the individuals, a crossover probability $p_c=0.8$, and a mutation probability $p_m = 0.05$.
As for the parameters of the GA, such as population size, crossover rate, and mutation rate, there is no universal criterion for their selection. These choices depend on factors like the nature of the problem, available computational resources, and the desired solution accuracy.
For instance, in \cite[Table I]{Hassanat_2019}, various percentages used in GA-based approaches across different studies are summarized.
In practice, many combinations of these parameters may lead to acceptable suboptimal solutions, or even a near-optimal solution.
The population size was chosen with respect to the computational capabilities of the hardware used for the optimization process\footnote{During the simulation work leading to the results shown here, different parameter combinations have been evaluated with many of them leading to very similar results. Study of the parameters is not presented in the manuscript due to length constrains.}. The complete parametrization is summarized in Table~\ref{tab:CGA-params}.

For the proposed \gls{PDGA}-\gls{ASW} algorithm, we adopt a maximum number of generations $G=500$ and an initial population size of $N_{\text{eval}}=10$, which also corresponds to the number of newly evaluated individuals per iteration.
In the considered \gls{CF-mMIMO} scenario, evaluating the fitness functions requires system-level simulations to compute both spectral and energy efficiency metrics, which makes each evaluation computationally costly.
When dealing with expensive multi-objective optimization problems, evolutionary algorithms are commonly designed to operate under a limited evaluation budget, where reducing the population size becomes an effective strategy to limit the number of costly fitness evaluations while maintaining adequate exploration of the solution space~\cite{Chugh2019}.
This design principle is consistent with micro-genetic and archive-based evolutionary algorithms, where very small populations combined with elitist or archive mechanisms have been shown to achieve competitive approximations of the Pareto front with significantly fewer evaluations~\cite{Coello2001,Tiwari2011}.
To compensate for the reduced population size, the number of generations is increased so that the algorithm can progressively refine the approximation of the Pareto front.
The remaining parameters mainly control the parent-selection pressure and the mutation operators used to explore alternative activation patterns.
The complete parametrization of the \gls{PDGA}-\gls{ASW} algorithm is summarized in Table~\ref{tab:PDGA-params}.

Since the primary optimization objective in the considered problem is energy efficiency, the evolutionary search is intentionally biased toward energy-efficient configurations.
The parent-selection and mutation mechanisms therefore introduce a controlled bias toward solutions with high energy efficiency, which accelerates the discovery of activation patterns that reduce the overall network power consumption.
In particular, the mutation operators that deactivate one or several \glspl{AP} promote the exploration of sparse activation patterns, which are typically associated with higher energy efficiency.
Consequently, the algorithm tends to identify high-energy-efficiency configurations faster than intermediate trade-off solutions.
To prevent the search from collapsing toward overly sparse activation patterns and to preserve the high spectral-efficiency region of the Pareto front, the all-APs-ON configuration is explicitly maintained within the population throughout the optimization process.

\subsection{PERFORMANCE EVALUATION}

\begin{figure*}[!t]
    \centering
    \subfloat[$\Upsilon$ vs. $L_A$, CB]{%
        \includegraphics[height=6cm]{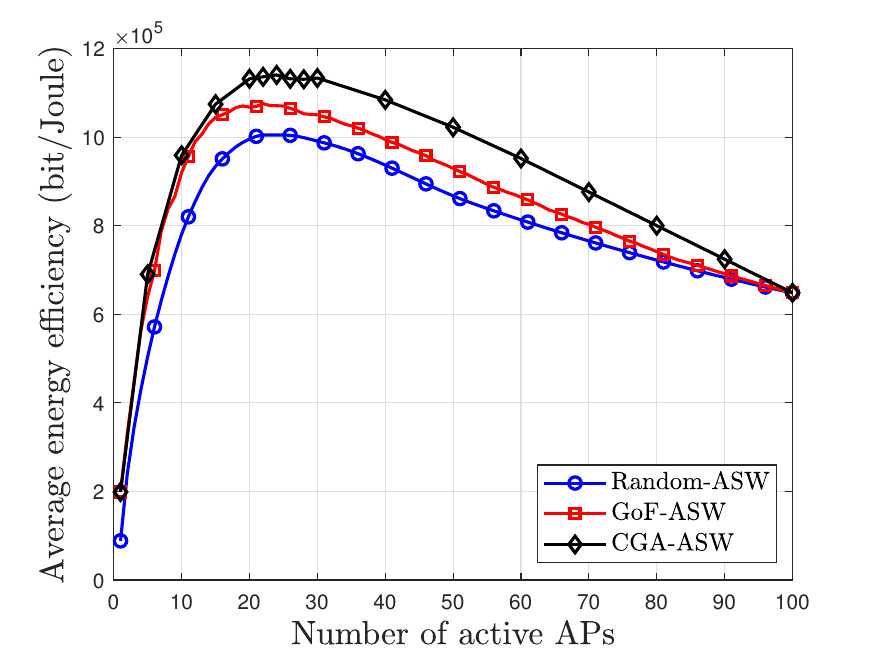}%
        \label{fig:Ee_vs_LA_MR}%
    }
    \hfill
    \subfloat[$\Upsilon$ vs. $L_A$, MMSE]{%
        \includegraphics[height=6cm]{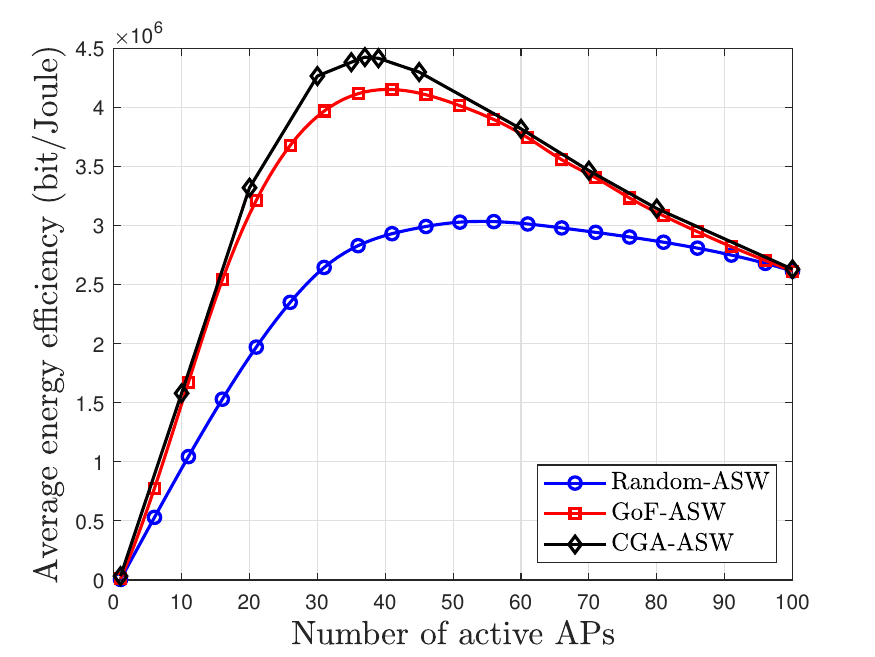}%
        \label{fig:Ee_vs_LA_MMSE}%
    }

    \par\vspace{1em}

    \subfloat[$\Xi$ vs. $L_A$, CB]{%
        \includegraphics[height=6cm]{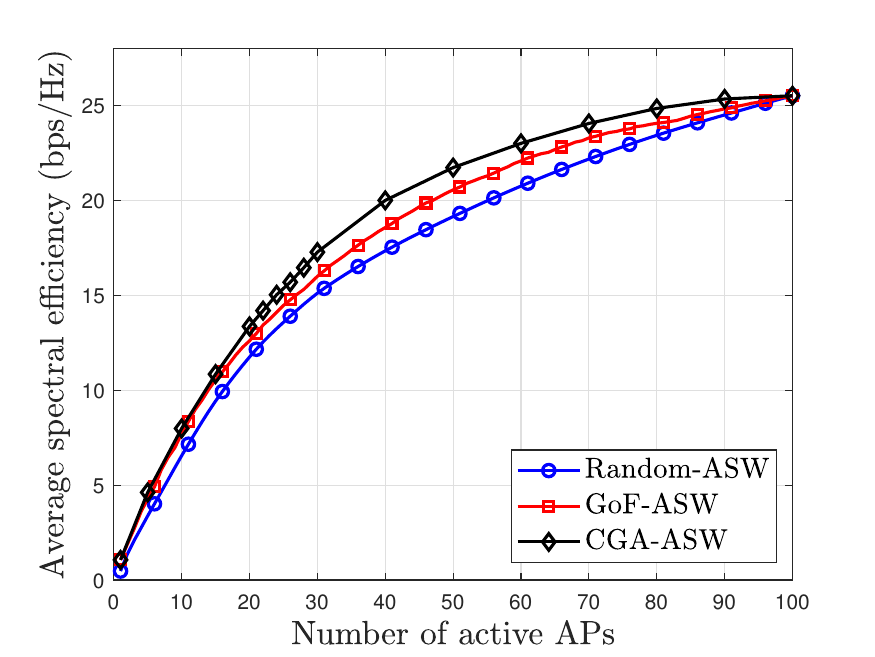}%
        \label{fig:Se_vs_LA_MR}%
    }
    \hfill
    \subfloat[$\Xi$ vs. $L_A$, MMSE]{%
        \includegraphics[height=6cm]{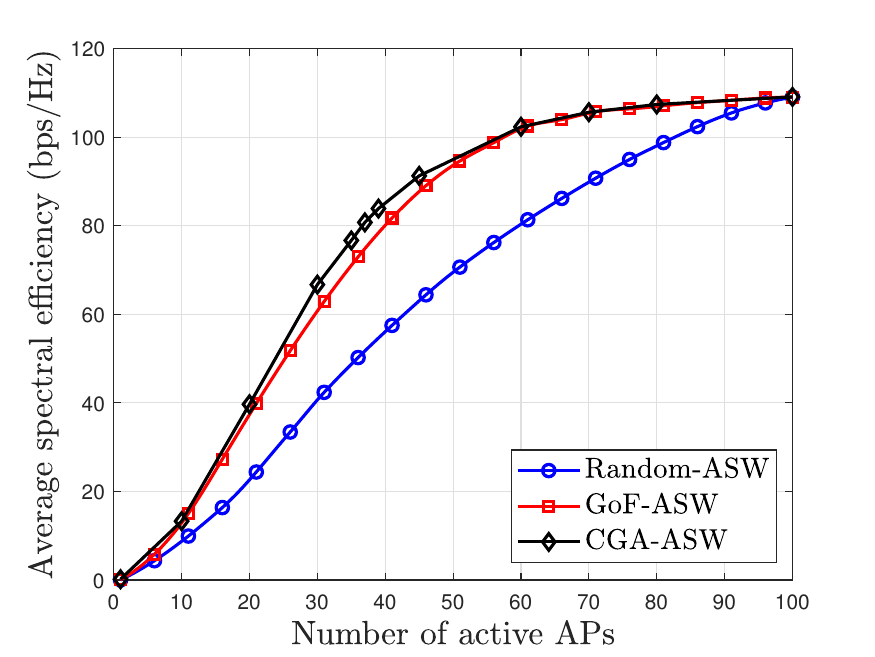}%
        \label{fig:Se_vs_LA_MMSE}%
    }

    \caption{\small Impact of the CGA-\gls{ASW} technique on the average \gls{EE} and \gls{SE} as a function of the number of active \glspl{AP}.}
    \label{fig:CGA_results}
\end{figure*}

Figure \ref{fig:CGA_results} presents the average energy and spectral efficiencies as a function of the number of active \glspl{AP} when considering the benchmark \gls{GoF}-based algorithm and the proposed CGA-\gls{ASW} strategy, for both \gls{CB} and \gls{MMSE} precoding schemes.
For comparison, a random selection strategy (denoted as Random-\gls{ASW}) is also included. 
This is an activity-count–based strategy, where a random subset $\mathcal{L}^A$ of active \glspl{AP} is chosen for each $L_A$ value.
Figure \ref{fig:CGA_results} shows that the proposed CGA-\gls{ASW} approach and the benchmark \gls{GoF}-\gls{ASW} strategy  outperform Random-\gls{ASW}.
The random selection strategy lacks awareness of the spatial distribution of \glspl{MS}, effectively assuming a homogeneous distribution.
Results obtained using the Random-\gls{ASW} strategy can serve as a lower-bound in terms of energy and spectral efficiency performance.
In contrast, both CGA-\gls{ASW} and \gls{GoF}-\gls{ASW} are designed for realistic inhomogeneous spatial traffic models.
Since the \gls{GoF}-\gls{ASW} strategy employs a greedy search algorithm to improve energy efficiency, it evaluates the solution space myopically, making decisions based solely on immediate energy savings by minimizing the discrepancy metric at each iteration, and therefore provides a poor exploration of the overall solution space.
The proposed CGA-\gls{ASW} strategy explores a broader solution space through population diversity and stochastic operators like mutation and crossover.
Consequently, CGA-\gls{ASW} outperforms \gls{GoF}-\gls{ASW}, as shown in Figure \ref{fig:CGA_results}, albeit at the cost of increasing the computational complexity as discussed in Section~\ref{sec:CGA_ASW}.


    

CGA-\gls{ASW} achieves a maximum energy efficiency of approximately $11.41 \times 10^5$ bit/Joule with $24$ active \glspl{AP} under  \gls{CB} precoding scheme, representing a \textbf{6.50\%} improvement over the greedy‑based \gls{GoF}-\gls{ASW}, which attains around $10.76 \times 10^5$ bit/Joule.
Similarly, with \gls{MMSE} processing, CGA-\gls{ASW} reaches about $44.29 \times 10^5$ bit/Joule with $37$ active \glspl{AP}, yielding a \textbf{7.12\%} gain relative to the $41.55 \times 10^5$ bit/Joule achieved by \gls{GoF}-\gls{ASW}.
The best near-optimal number of active \glspl{AP} for \gls{GoF}-\gls{ASW} is $22$ for \gls{CB}, and $40$ for \gls{MMSE}, showing a slight deviation from the best near-optimal values obtained with CGA-\gls{ASW} for each precoding scheme.

When comparing \gls{CB} and \gls{MMSE} precoding schemes, \gls{MMSE} clearly offers superior performance due to its active interference suppression, albeit at the cost of higher computational complexity. 
In contrast, the low-complexity \gls{CB} scheme disregards interference, resulting in lower performance bounds.
For a given \gls{ASW} technique, the best near-optimal number of active \glspl{AP} obtained when maximizing energy efficiency is consistently lower with \gls{CB} than with \gls{MMSE}. 
Since \gls{CB} does not manage interference, activating additional \glspl{AP} beyond the best near-optimal point obtained yields only marginal gains in spectral efficiency, insufficient to offset the increased energy consumption. 
In comparison, \gls{MMSE}, by effectively suppressing interference, can accommodate a larger number of active \glspl{AP} before reaching its best near-optimal configuration.

\begin{figure*}[!t]
    \centering
    \subfloat[CB]{%
        \includegraphics[height=6cm]{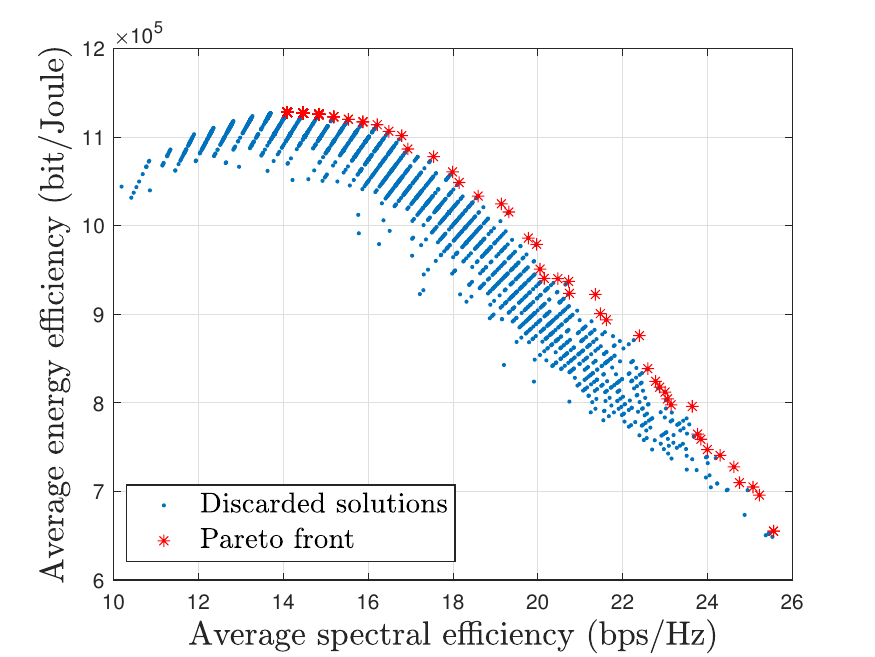}%
        \label{fig:PDGA_results_MR}%
    }
    \hfill
    \subfloat[MMSE]{%
        \includegraphics[height=6cm]{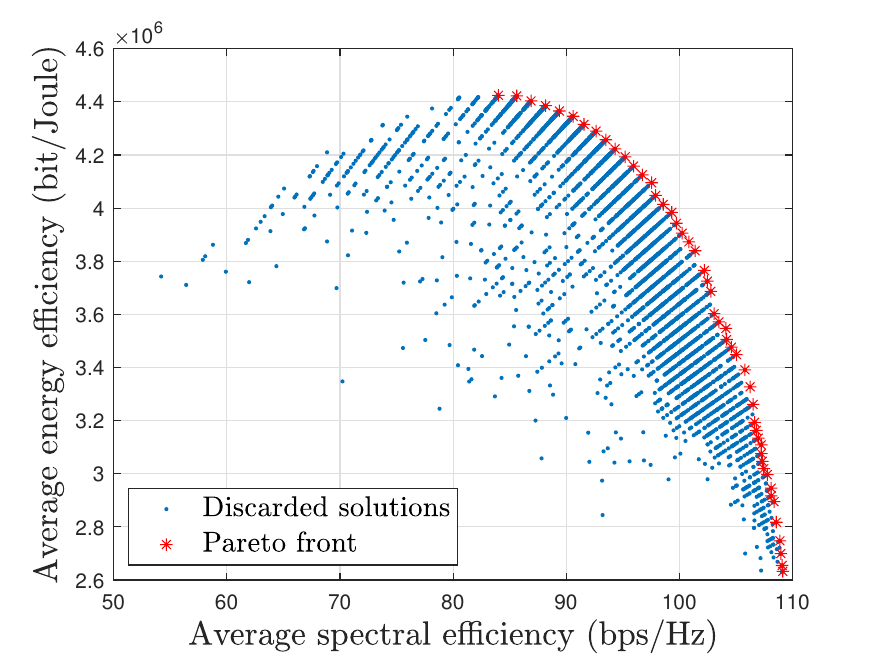}%
        \label{fig:PDGA_results_MMSE}%
    }
    
    \caption{\small Combined performance of the \gls{PDGA}-\gls{ASW} strategy solutions.}
    \label{fig:PDGA_results}
\end{figure*}

\begin{figure*}[h!]
    \centering
    \subfloat[$\max\Upsilon$ and $\text{optimal }L_A$ vs. $\Xi_{\min}$, CB]{%
        \includegraphics[height=6cm]{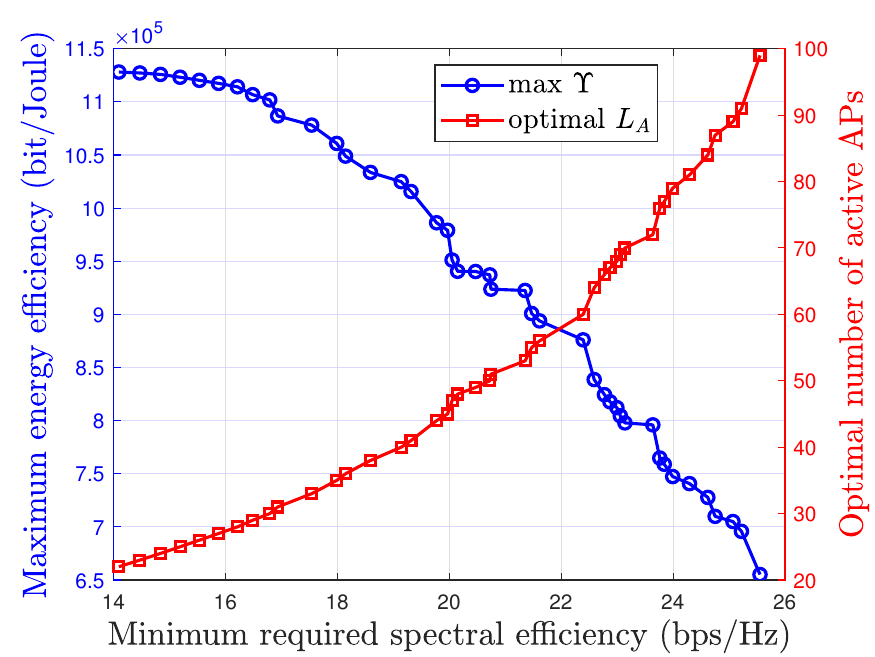}%
        \label{fig:EE_LA_SE_results_MR}%
    }
    \hfill
    \subfloat[$\max\Upsilon$ and $\text{optimal }L_A$ vs. $\Xi_{\min}$, MMSE]{%
        \includegraphics[height=6cm]{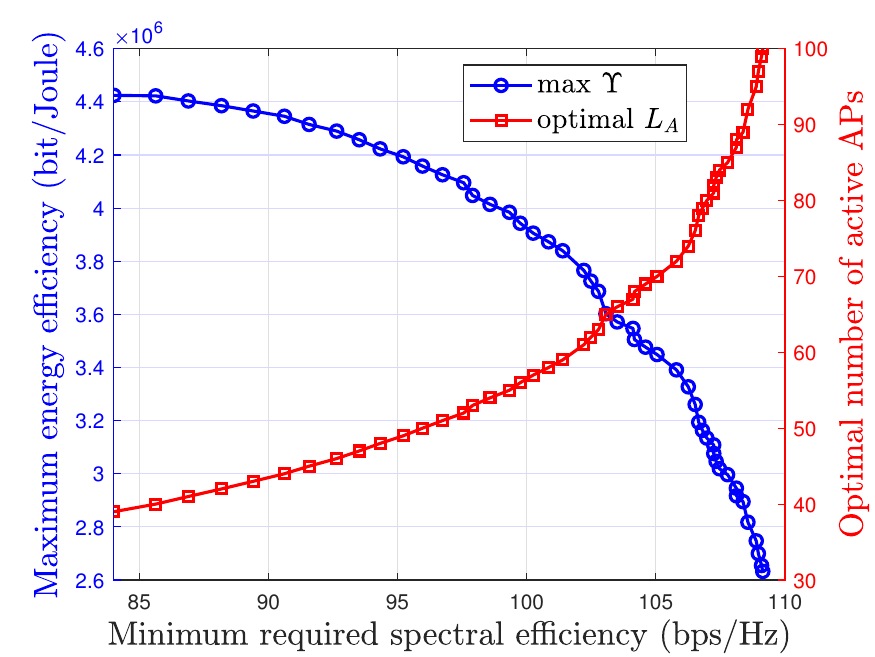}%
        \label{fig:EE_LA_SE_results_MMSE}%
    }
    \caption{\small EE operating points for different SE requirements.}
    \label{fig:operating_points}
\end{figure*}

To evaluate the performance of the proposed \gls{PDGA}-\gls{ASW} technique, Fig.~\ref{fig:PDGA_results} presents the average energy efficiency versus spectral efficiency.
The Pareto front solutions obtained using the \gls{PDGA}-\gls{ASW} illustrate the trade-off between energy and spectral efficiency, highlighting the balance achieved when improving one metric without excessively degrading the other.
Fig.~\ref{fig:PDGA_results} also reveals large regions of dominated solutions in the energy–spectral efficiency plane, emphasizing the advantage of using more powerful \gls{ASW} strategies over random selection.
Notably, the region of dominated solutions is larger when using \gls{MMSE} precoding compared to \gls{CB}.
Furthermore, the Pareto front obtained with \gls{MMSE} includes both higher energy efficiency and superior spectral efficiency values.
This is because the \gls{MMSE} processing enables \gls{PDGA}-\gls{ASW} to find solutions that better suppress interference and eliminate many suboptimal configurations.
While the Pareto front provides a comprehensive characterization of the achievable trade-off between \gls{EE} and \gls{SE}, its practical usefulness ultimately lies in enabling the selection of a suitable operating point according to system-level requirements. 
In practical \gls{CF-mMIMO} deployments, the network operation is typically constrained by \gls{QoS} requirements, which can be naturally expressed in terms of a minimum network \gls{SE}.
A representative operating point can be selected from the Pareto front by solving: 
$\max_{\mathcal{L}_A \subseteq \mathcal{L}} \Upsilon(\mathcal{L}_A) \quad \text{s.t.} \quad \Xi(\mathcal{L}_A) \geq \Xi_{\min}$, 
where $\Xi_{\min}$ denotes the minimum network \gls{SE} required by the system.
This criterion allows selecting the most energy-efficient AP activation pattern while guaranteeing a target level of throughput. 
From the Pareto front shown in Fig.~\ref{fig:PDGA_results}, it can be observed that operating points beyond a certain region yield only marginal gains in \gls{SE} at the cost of a significant degradation in \gls{EE}.
Therefore, selecting a point close to the onset of this region (often referred to as the ``knee'' of the Pareto curve) provides a balanced trade-off between both metrics.
This emphasizes the relevance of the proposed \gls{PDGA}-based framework, which enables identifying such regions and selecting operating points that would be difficult to determine using single-objective or greedy approaches.


\begin{figure*}[!t]
    \centering
    \subfloat[$\Upsilon$ vs. $L_A$, CB]{%
        \includegraphics[height=6cm]{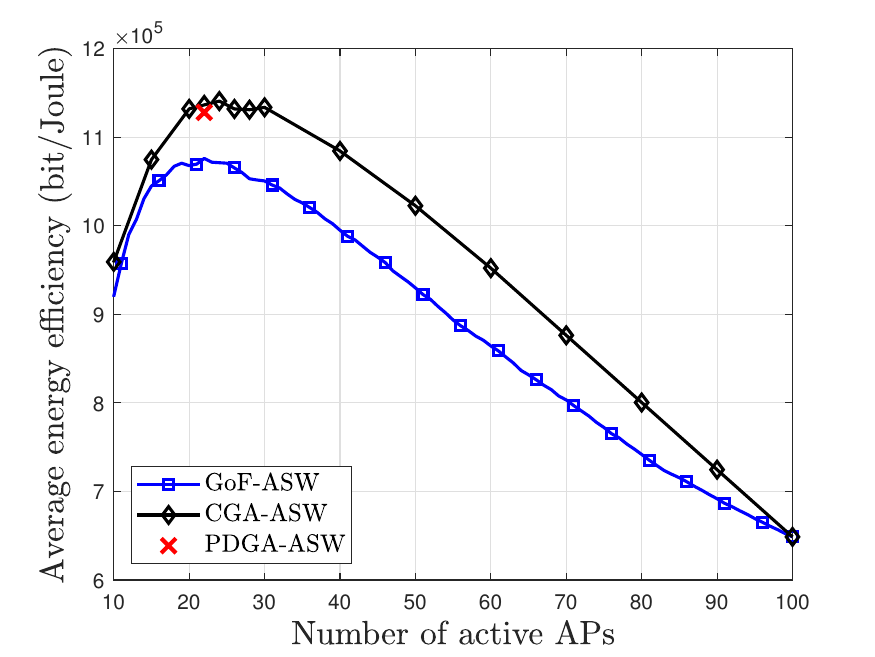}%
        \label{fig:impact_results_MR}%
    }
    \hfill
    \subfloat[$\Upsilon$ vs. $L_A$, MMSE]{%
        \includegraphics[height=6cm]{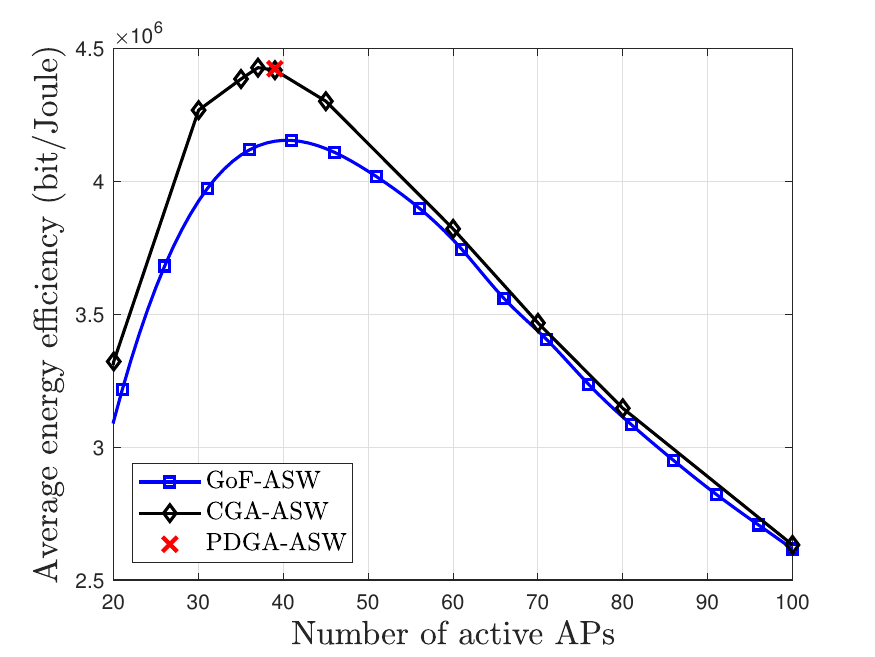}%
        \label{fig:impact_results_MMSE}%
    }

    \caption{\small Impact of \gls{CGA}-\gls{ASW} and \gls{PDGA}-\gls{ASW} strategies on the average \gls{EE} as a function of the number of active \glspl{AP}.}
    \label{fig:impact_results}
\end{figure*}

To further illustrate the decision-making capability enabled by the proposed PDGA-based framework, Fig.~\ref{fig:operating_points} analyzes the impact of different network \gls{SE} requirements for selected operating points. Specifically, we consider a set of $\Xi_{\min}$ obtained from the Pareto front under both \gls{CB} and \gls{MMSE} precoding schemes. 
For each value of $\Xi_{\min}$, the corresponding operating point is selected as the solution that maximizes \gls{EE} while satisfying the constraint \mbox{$\Xi(\mathcal{L}_A)\geq\Xi_{\min}$}. The results show that higher network \gls{SE} requirements lead to the activation of a larger number of \glspl{AP}, which increases the overall power consumption and reduces \gls{EE}. Conversely, relaxed \gls{SE} constraints enable significantly more energy-efficient configurations with fewer active \glspl{AP}. 
These results confirm that the proposed framework allows adapting the network operation to different \gls{QoS} regimes, highlighting its practical applicability for dynamic and heterogeneous \gls{CF-mMIMO} scenarios. Note that, as expected and in comparison to \gls{CB}, \gls{MMSE} leads to much larger \gls{EE} while supporting much more stringent {QoS} requirements.

Unlike the \gls{CGA}-\gls{ASW} proposal and the benchmark scheme, when using \gls{PDGA}-\gls{ASW}, an energy efficiency curve as a function of $L_A$ is not directly obtained.
Due to this, only the maximum energy efficiency point from Pareto front solutions is provided in Fig.~\ref{fig:impact_results}, which presents the average energy efficiency as a function of the number of active \glspl{AP} ($L_A$).
This figure highlights the impact of both \gls{CGA}-\gls{ASW} and \gls{PDGA}-\gls{ASW} in comparison to the benchmark \gls{GoF}-based strategy. 
First, it is evident that both proposed techniques outperform the benchmark in terms of energy efficiency. 
It is worth noting that the optimal solutions obtained using either \gls{CGA}-\gls{ASW} or \gls{PDGA}-\gls{ASW} are essentially equivalent, with minor differences arising from the reliance of the optimization procedure on Monte Carlo simulations.
Specifically, \gls{PDGA}-\gls{ASW} reaches approximately $11.28 \times 10^5$ bit/Joule with $22$ active \glspl{AP} under \gls{CB} precoding, and around $44.25 \times 10^5$ bit/Joule with $39$ active \glspl{AP} under \gls{MMSE} precoding.

\begin{figure*}[!t]
    \centering
    \subfloat[GoF-\gls{ASW}, $L_A = 22$, CB]{%
        \includegraphics[height=4.5cm]{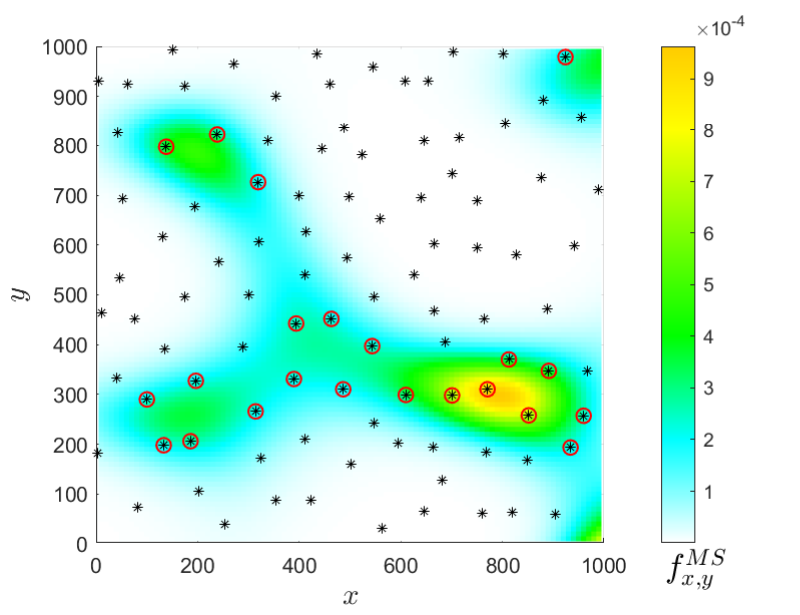}%
        \label{fig:GoF_MR_results}%
    }
    \hfill
    \subfloat[CGA-\gls{ASW}, $L_A = 22$, CB]{%
        \includegraphics[height=4.5cm]{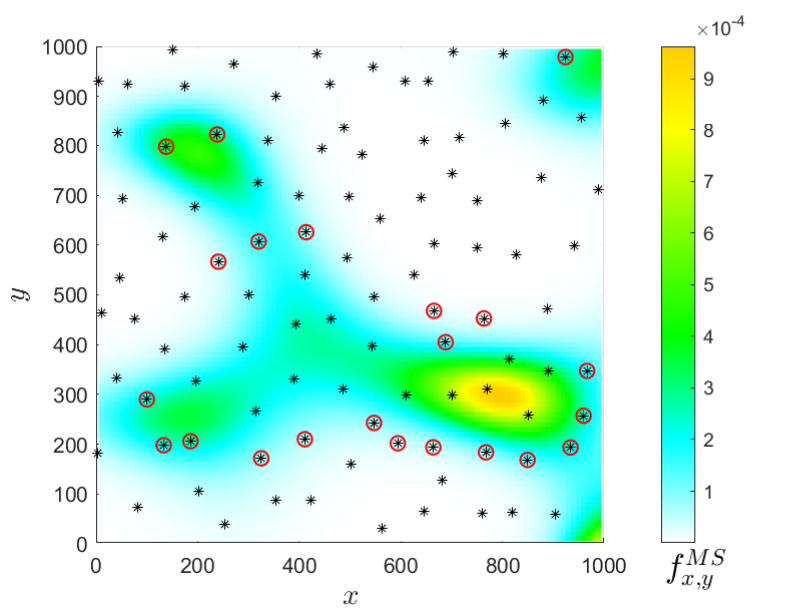}%
        \label{fig:CGA_MR_results}%
    }
    \hfill
    \subfloat[PDGA-\gls{ASW}, $L_A = 22$, CB]{%
        \includegraphics[height=4.5cm]{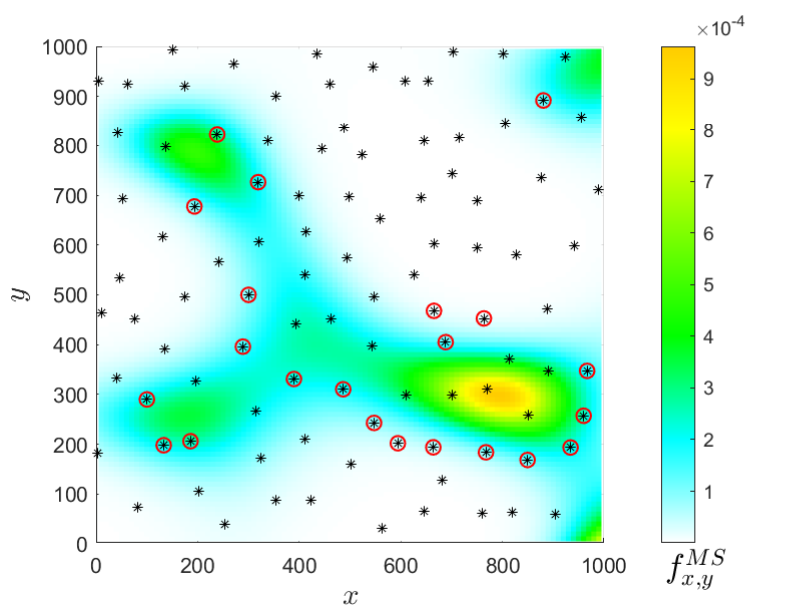}%
        \label{fig:PDGA_MR_results}%
    }

    \par\vspace{1em}

    \subfloat[GoF-\gls{ASW}, $L_A = 39$, MMSE]{%
        \includegraphics[height=4.5cm]{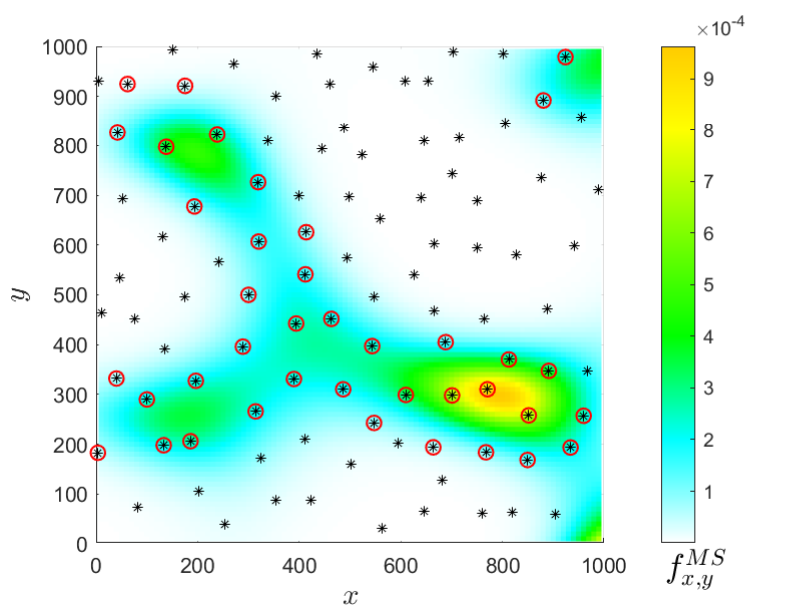}%
        \label{fig:GoF_MMSE_results}%
    }
    \hfill
    \subfloat[CGA-\gls{ASW}, $L_A = 39$, MMSE]{%
        \includegraphics[height=4.5cm]{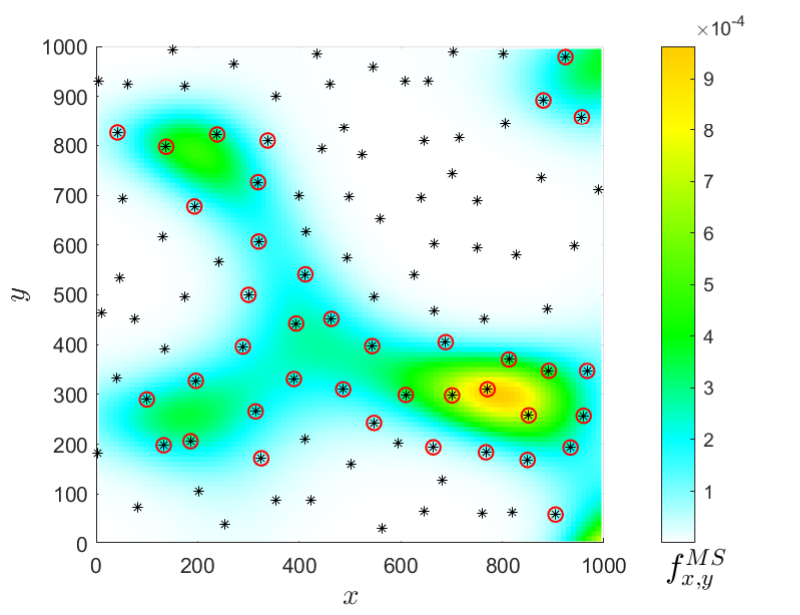}%
        \label{fig:CGA_MMSE_results}%
    }
    \hfill
    \subfloat[PDGA-\gls{ASW}, $L_A = 39$, MMSE]{%
        \includegraphics[height=4.5cm]{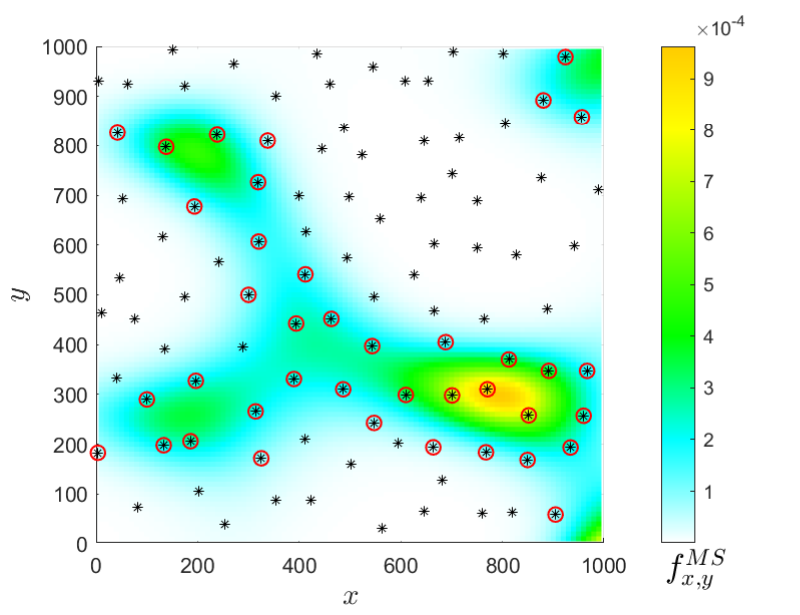}%
        \label{fig:PDGA_MMSE_results}%
    }

    \caption{\small Optimal/near-optimal solutions of GoF-\gls{ASW}, \gls{CGA}-\gls{ASW}, and \gls{PDGA}-\gls{ASW} techniques, for both \gls{CB} and \gls{MMSE} precoding schemes. Black asterisks represent the \glspl{AP} deployment, where the active ones are marked with a red circle. The \gls{MS} log-normal- based distribution is mapped from Fig.~\ref{fig:MS_pdf} in terms of $f^{\gls{MS}}_{x,y}$.}
    \label{fig:sols_results}
\end{figure*}

Figure~\ref{fig:sols_results} presents the near-optimal solutions obtained by the  GoF-\gls{ASW}, \gls{CGA}-\gls{ASW}, and \gls{PDGA}-\gls{ASW} techniques under both \gls{CB} and \gls{MMSE} precoding schemes.
The color map illustrates the distribution of \glspl{MS} in the coverage area.
For comparison purposes, a common near-optimal number of active \glspl{AP} has been selected for each precoding scheme across all techniques.
As illustrated in Figures~\ref{fig:GoF_MR_results} and~\ref{fig:GoF_MMSE_results}, the GoF-\gls{ASW} strategy effectively aligns the spatial distribution of active \glspl{AP} with that of the \glspl{MS} in the network.
Conversely, the proposed strategies under \gls{CB} precoding, shown in Figures~\ref{fig:CGA_MR_results} and ~\ref{fig:PDGA_MR_results}, aim to balance service across all \glspl{MS} by activating \glspl{AP} that maximize average energy efficiency.
As a result, these solutions avoid activating \glspl{AP} that are located too close to \glspl{MS}, in order to reduce interference, which is not mitigated by the \gls{CB} precoding scheme.
In contrast, under \gls{MMSE} precoding, which can effectively manage interference, the optimal or near-optimal solutions obtained by \gls{CGA}-\gls{ASW} and \gls{PDGA}-\gls{ASW} (Figures~\ref{fig:CGA_MMSE_results} and ~\ref{fig:PDGA_MMSE_results}, respectively) closely resemble the benchmark solution shown in Fig.~\ref{fig:GoF_MMSE_results}.
These results are in line with those shown in Figures~\ref{fig:Ee_vs_LA_MMSE} and~\ref{fig:Se_vs_LA_MMSE}, where the optimal solutions obtained by the proposed \gls{ASW} and \gls{GoF}-\gls{ASW} are very similar when \gls{MMSE} precoding is used.
This similarity arises because \gls{GoF}-\gls{ASW} activates \glspl{AP} based on the spatial distribution of \glspl{MS}, while \gls{CGA}-\gls{ASW} and \gls{PDGA}-\gls{ASW}, under \gls{MMSE}, select \glspl{AP} that are physically closest to \glspl{MS}, an approach that effectively approximates the \glspl{MS} distribution.

\begin{figure*}[!t]
    \centering
    \subfloat[Single-hotspot distribution]{%
        \includegraphics[height=4.4cm]{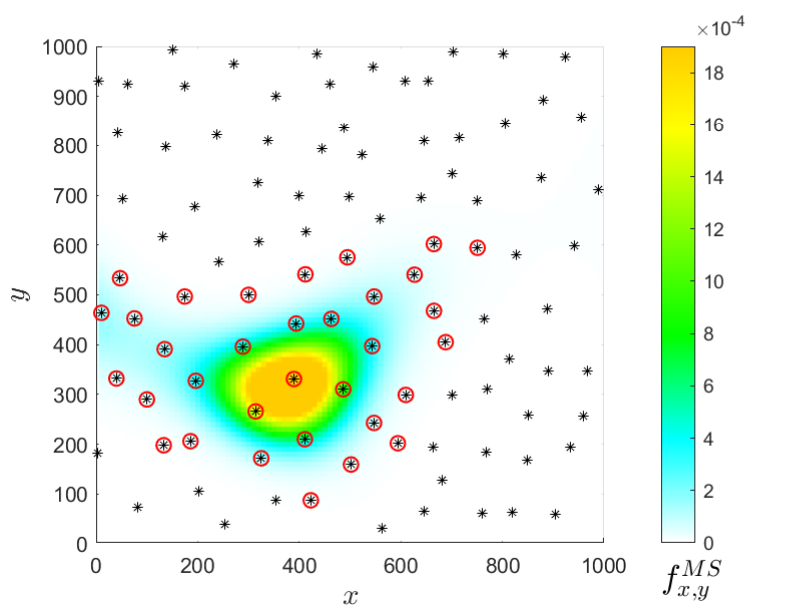}%
        \label{fig:PDGA_MMSE_Hotspot}
    }
    \hfill
    \subfloat[Multiple-hotspot distribution]{%
        \includegraphics[height=4.4cm]{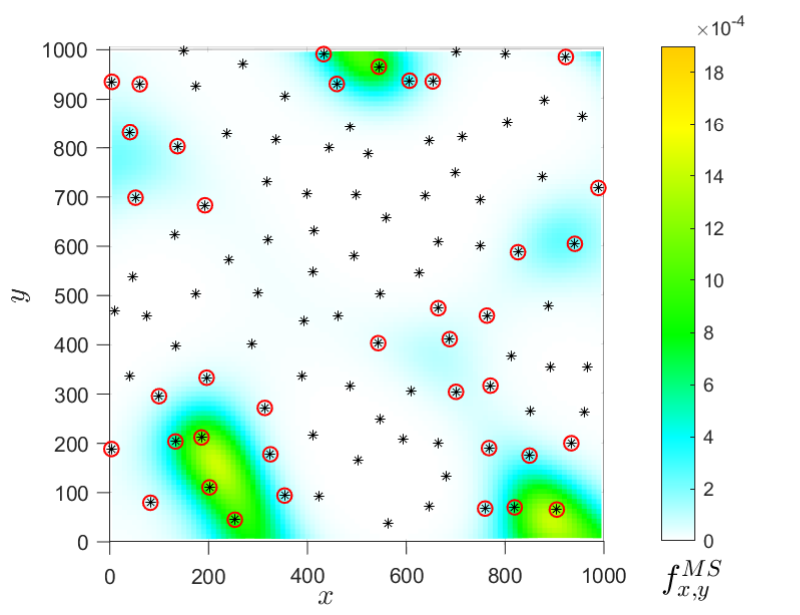}%
        \label{fig:PDGA_MMSE_3Hotspot}
    }
    \hfill
    \subfloat[Uniform distribution]{%
        \includegraphics[height=4.4cm]{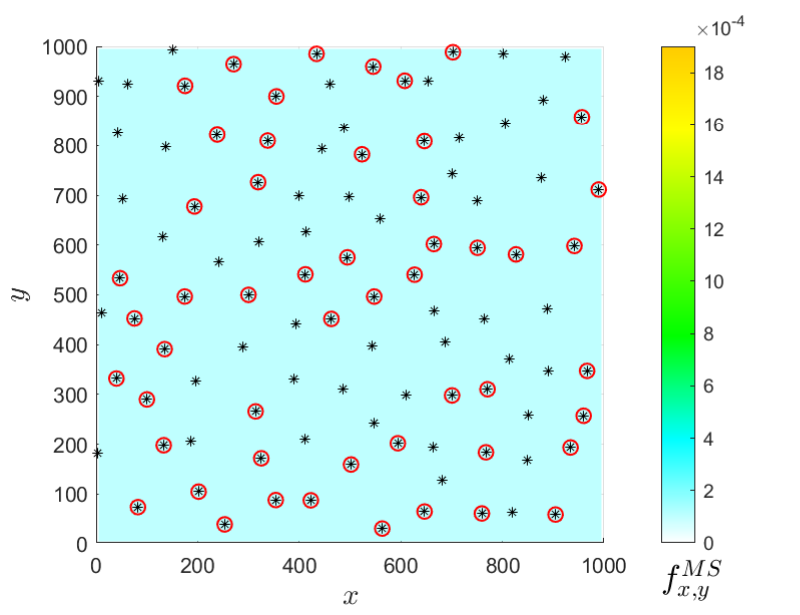}%
        \label{fig:PDGA_MMSE_Uniform}
    }
    
    \caption{\small Optimal/near-optimal solutions for different spatial \gls{MS} distributions using \gls{PDGA}-\gls{ASW} strategy with \gls{MMSE} precoding. Black asterisks represent the \glspl{AP} deployment, where the active ones are marked with a red circle. Each distribution is mapped in terms of the probability density function $f^{\gls{MS}}_{x,y}$.}
    \label{fig:more_sols_results}
\end{figure*}

To validate that the proposed framework is deployment-agnostic and robust across different scenarios, different spatial \gls{MS} distributions are analyzed in Fig.~\ref{fig:more_sols_results}, which depicts the corresponding near-optimal solutions obtained by the proposed \gls{PDGA}-\gls{ASW} algorithm combined with \gls{MMSE} precoding scheme.
The configurations result from optimizing the ON/OFF switching of \glspl{AP} to maximize the network \gls{EE}.
Specifically, the following cases are considered: (i) a single-hotspot distribution (Fig.~\ref{fig:PDGA_MMSE_Hotspot}), (ii) a multiple-hotspot distribution with three main activity hotspots (Fig.~\ref{fig:PDGA_MMSE_3Hotspot}), and (iii) a uniform distribution (Fig.~\ref{fig:PDGA_MMSE_Uniform}).
Qualitatively, the resulting active \gls{AP} configurations closely match the underlying spatial \gls{MS} distributions, consistent with the previous results obtained when using \gls{MMSE} (see Figures~\ref{fig:GoF_MMSE_results}--\ref{fig:PDGA_MMSE_results}).
Quantitatively, the single-hotspot case achieves a maximum {EE} of $3.36 \times 10^6$ bit/Joule with only 33 active APs.
In contrast, a more spatially distributed scenario with three main hotspots attains a lower maximum \gls{EE} of $2.73 \times 10^6$ bit/Joule with 38 active APs, since the optimization leads to a more spatially expansive AP activation to cover multiple high-traffic regions.
Subsequently, a homogeneous scenario with a uniform \gls{MS} distribution yields an optimized configuration with 52 uniformly distributed active \glspl{AP}, achieving the lowest maximum \gls{EE} of $2.44 \times 10^6$ bit/Joule for the considered \gls{CF-mMIMO} network.
Notably, when the spatial \gls{MS} distribution is uniform, the solution can be simplified to determining the number of active \glspl{AP}, provided that the active \glspl{AP} are distributed as uniformly as possible.
Overall, the optimal number of active \glspl{AP} is jointly determined by the spatial \gls{MS} distribution, the interference level, and the adopted interference management strategy in \gls{CF-mMIMO} networks.

\subsection{COMPUTATIONAL COST AND IMPLEMENTATION ASPECTS}

This section analyzes the computational cost of the proposed \gls{AP} ON/OFF optimization framework and discusses the hardware requirements needed to apply it in practical network-planning scenarios. The main objective is to show that the proposed methods are computationally feasible when applied to \gls{MS} distributions that evolve slowly, typically over time scales on the order of hours.
This assumption is consistent with the type of traffic variations targeted by energy-saving mechanisms in mobile networks. Network load exhibits strong daily and hourly patterns, with clear differences between peak and off-peak periods. Moreover, industrial energy-saving mechanisms are commonly designed around the dynamic activation, deactivation, or adaptation of radio resources according to traffic conditions and load patterns, especially during low-load periods. Therefore, the relevant time scale for the proposed \gls{AP} ON/OFF optimization is not the short-term traffic fluctuation at the level of seconds or minutes, but the slower evolution of representative spatial \gls{MS} distributions.

All execution-time measurements were obtained using \gls{CPU}-based workstations and the software Matlab R2024b, representative of the computational environment used for the optimization experiments and evaluation procedures.
The reference hardware platform consists of Intel Core i9-class \glspl{CPU} with 32 GB RAM, defining the effective memory constraint considered in the computational assessment.
Accordingly, the reported execution times correspond to a moderately constrained workstation environment rather than a high-performance computing cluster, reflecting the objective of assessing whether the proposed optimization can be deployed with practical resources, e.g., using a limited number of similar workstations or server nodes.

\vspace{0.2cm}
{\bf Average fitness execution time per AP configuration:}

\begin{figure}[t!]
    \centering
    \includegraphics[width=\linewidth]{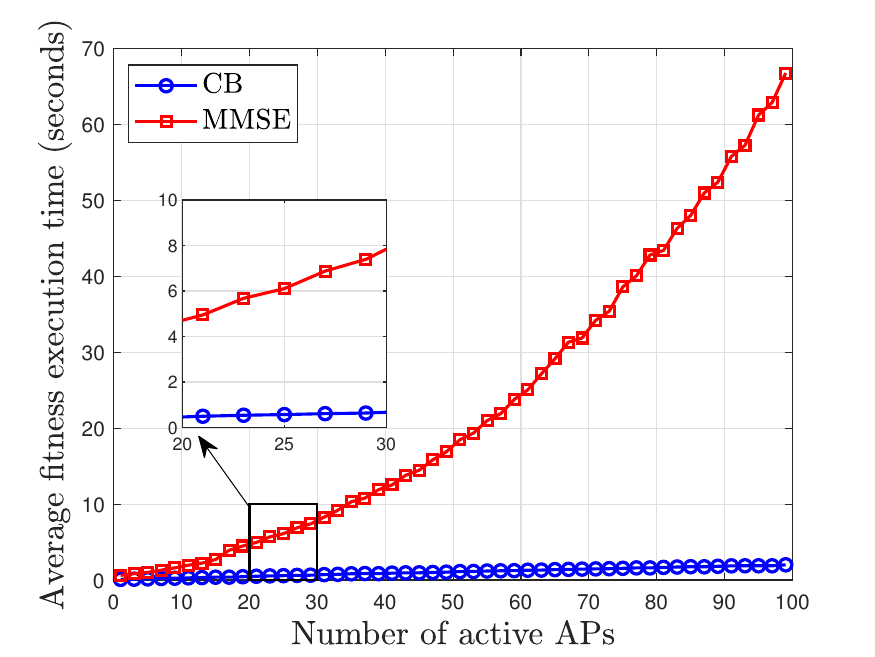}
    \caption{\small Execution time required to evaluate one \gls{AP} ON/OFF configuration as a function of $L_A$.}
    \label{fig:average_time_per_configuration}
\end{figure}

The evaluation of a candidate \gls{AP} ON/OFF configuration requires computing the resulting system performance, including \gls{SE} and \gls{EE} associated with the selected set of active \glspl{AP}.
The cost of one evaluation is not constant, since it depends on the number of active \glspl{AP} $L_A$.
To characterize this dependency, the execution time of one configuration evaluation was measured for different values of $L_A$.
Let $\bar{t}(L_A)$ denote the average time, in seconds, required to evaluate one candidate configuration with $L_A$ active \glspl{AP}.
Figure~\ref{fig:average_time_per_configuration} shows the measured average execution time as a function of the number of active \glspl{AP} (i.e., $\bar{t}(L_A)$ vs. $L_A$).
The curve confirms that the evaluation cost increases with the number of active \glspl{AP}.
Therefore, a fair comparison between optimization strategies should account not only for the number of fitness evaluations, but also for the AP-cardinality (i.e., $L_A$ value) of those evaluations.

\vspace{0.2cm}
{\bf Execution-time estimation for \gls{CGA}:}

When considering the \gls{CGA}-based approach, the number of active \glspl{AP} is fixed during each optimization run. Therefore, identifying the \gls{AP}-cardinality that maximizes \gls{EE} requires executing \gls{CGA} for several values of $L_A$. Since the \gls{EE} curve as a function of $L_A$ has a single maximum (see Figures~\ref{fig:Ee_vs_LA_MR} and~\ref{fig:Ee_vs_LA_MMSE}), the cardinality search can be performed through a discrete interval-refinement procedure (or by using the bisection method~\cite{boyd2004convex}) instead of an exhaustive sweep over all possible values.
Therefore, the \gls{CGA}-based search requires independent optimizations for the set $\mathcal{S}_{\mathrm{CGA}}$ of $L_A$ representative values.
For example, under \gls{MMSE} processing, the procedure starts by evaluating three coarse \gls{AP}-cardinality values, $L_A = 25, 50, 75$.
In the considered case, the high-cardinality region did not provide the largest \gls{EE}, so the search is refined in the interval between $L_A=25$ and $L_A=50$.
The intermediate value $L_A = 37$ was then evaluated.
Since this value provided a higher \gls{EE} than the previously evaluated cardinalities, the center of the search is updated to $L_A=37$.
The search was then locally refined by evaluating $L_A = 43, 40, 39$.
Finally, the neighboring cardinalities $L_A = 38, 36$, are evaluated to strictly verify that $L_A=37$ is the discrete maximum of the unimodal curve.
Therefore, the \gls{CGA}-based search required independent optimizations for the set $\mathcal{L}_{A_\mathrm{CGA}}=\{25,50,75,37,43,40,39,38,36\}$ of $L_A$ values, under \gls{MMSE} processing.
Additionally, each \gls{CGA} execution for a fixed $L_A$ performs a total of $N_{\text{eval}} + G \cdot N_{\text{eval}}$ fitness function evaluations (i.e., $5050$ fitness evaluations using the data in Table~\ref{tab:CGA-params}).
Hence, the total execution time $T_{\mathrm{CGA}}$ (in hours) of the CGA-based search is estimated as
\setcounter{equation}{31}
\begin{equation}
T_{\mathrm{CGA}}
=
\frac{N_{\text{eval}} + G \cdot N_{\text{eval}}}{60 \cdot 60}
\sum_{L_A \in \mathcal{L}_{A_\mathrm{CGA}}}
\bar{t}(L_A),
\end{equation}
where the factor $(60 \cdot 60)$ converts the result from seconds to hours.

\vspace{0.2cm}
{\bf Execution-time estimation for \gls{PDGA}:}

For the \gls{PDGA}-based approach, the total execution time is estimated by weighting the measured average evaluation time for each \gls{AP}-cardinality by the number of candidate configurations evaluated at that cardinality. Therefore, the total time $T_{\mathrm{PDGA}}$ (in hours) can be written as

\begin{equation}
T_{\mathrm{PDGA}}
=
\frac{1}{60 \cdot 60}
\sum_{\forall L_A}
\bar{t}(L_A)n_{\mathrm{PDGA}}(L_A),
\end{equation}

where $n_{\mathrm{PDGA}}(L_A)$ is the number of \gls{PDGA} evaluations performed with \(L_A\) active \glspl{AP}.
The values of $n_{\mathrm{PDGA}}(L_A)$, $\forall L_A$, are obtained by simulation.

\vspace{0.2cm}
{\bf Feasibility under slowly varying \gls{MS} distributions:}

The practical applicability of the proposed method relies on the fact that the relevant spatial \gls{MS} density distributions evolve more slowly than instantaneous traffic fluctuations. In operational networks, \gls{MS} traffic and load conditions exhibit strong temporal regularities associated with human activity patterns, such as residential, business, transport, and entertainment areas. These patterns typically evolve over hourly and daily time scales rather than at the level of a few minutes.
Accordingly, the proposed method should be interpreted as a planning and slow-timescale adaptation tool. The optimization is intended to be executed for representative \gls{MS} distributions, for example those associated with different periods of the day or with specific spatial traffic regimes. Once optimized \gls{AP} activation patterns are available, the operational system can select or update the appropriate configuration according to the estimated current \gls{MS} density regime, rather than solving a full combinatorial optimization problem from scratch for every short-term traffic fluctuation.
This interpretation is also aligned with industrial energy-saving approaches.
In Open virtualized Radio Access Network (vRAN) and 5G Radio Access Network (RAN) systems, energy-saving mechanisms commonly exploit the possibility of deactivating or adapting capacity resources during low-load periods, while maintaining coverage through the remaining active resources.
Such mechanisms are naturally associated with traffic-load regimes and off-peak periods, not with continuous minute-by-minute topology re-optimization.

\begin{table}[t!]
\centering

\setlength{\tabcolsep}{3pt}
\renewcommand{\arraystretch}{1.0}

\caption{\small Empirical runtime decomposition for the AP-switching methods}
\label{tab:runtime_decomposition}

\begin{tabular}{l l c c c c}
\hline
\textbf{Method}
& \textbf{Scheme}
& \begin{tabular}{c}\textbf{Fitness}\\\textbf{evaluation}\end{tabular}
& \begin{tabular}{c}\textbf{Search/}\\\textbf{ordering}\end{tabular}
& \begin{tabular}{c}\textbf{Total}\\\textbf{runtime}\end{tabular}
& \begin{tabular}{c}\textbf{Overhead}\\\textbf{share}\end{tabular} \\
\hline
PDGA & CB   & 1.44 h    & 14.11 s   & $\approx$ 1.44 h   & 0.27\% \\
PDGA & MMSE & 26.70 h   & 17.56 s   & $\approx$ 26.70 h  & 0.02\% \\
CGA  & CB   & 11.64 h   & 1.10 s    & $\approx$ 11.64 h  & 0.003\% \\
CGA  & MMSE & 184.40 h  & 1.10 s    & $\approx$ 184.40 h & 0.0002\% \\
GoF  & CB   & 1.78 min  & 16.30 min & 18.08 min          & 90.15\% \\
GoF  & MMSE & 38.60 min & 16.30 min & 54.90 min          & 29.69\% \\
\hline
\end{tabular}
\end{table}

\begin{table}[t!]
\centering
\caption{\small Approximate number of equivalent 32 GB workstation nodes required to complete one optimization within different MS distribution update periods.}
\label{tab:hardware_requirements}
\begin{tabular}{c c c c}
\hline
\textbf{Processing} & \textbf{Update period} & \textbf{CGA} & \textbf{PDGA} \\
 & \(\Delta t\) & $N_{\mathrm{eq,CGA}}$ & $N_{\mathrm{eq,PDGA}}$ \\
\hline
CB & 30 min & 46.6 & 2.9 \\
CB & 40 min & 34.9 & 2.2 \\
CB & 50 min & 23.3 & 1.7 \\
CB & 60 min & 11.6 & 1.4 \\
\hline
MMSE & 4 h  & 46.1 & 6.7 \\
MMSE & 8 h  & 23.1 & 3.3 \\
MMSE & 12 h & 15.4 & 2.2 \\
MMSE & 24 h & 7.7 & 1.1 \\
\hline
\end{tabular}
\end{table}

Table~\ref{tab:runtime_decomposition} presents the empirical runtime decomposition of the AP-switching methods.
The \gls{CGA} and \gls{PDGA} approaches perform \gls{EE}-driven optimization by repeatedly evaluating the physical-layer fitness function under \gls{CB} or \gls{MMSE} processing.
In contrast, GoF-\gls{ASW} is included as a spatial \gls{MS} distribution-based baseline: it constructs the \gls{AP} activation sequence using a goodness-of-fit criterion and evaluates the resulting configurations a posteriori.
As observed in Table~\ref{tab:runtime_decomposition}, the \gls{MMSE} scheme requires a longer runtime, which reflects the cost associated with its superior performance in comparison with \gls{CB}.
Similarly, the \gls{GA}-based approaches exhibit higher execution times compared to the GoF-\gls{ASW} baseline, in exchange for improved energy-efficiency optimization.
Remarkably, the \gls{PDGA}-\gls{ASW} achieves lower runtime values compared to the \gls{CGA}-\gls{ASW}, with the same exploration of the energy-{spectral} efficiency trade-off (see Fig.~\ref{fig:impact_results}).
This efficiency gain highlights one of the main advantages of the proposed \gls{PDGA} approach.
In line with the dominance analyses in Section~\ref{sec:GA_ASW}, fitness evaluation across generations and individuals dominates the overall computational cost, with runtimes significantly exceeding those of search and ordering operations. These results confirm that, although some terms scale rapidly with algorithm parameters (e.g., in \gls{PDGA}, Pareto front identification scales quadratically with \(N_{\text{eval}}\) and cubically with \(G\)), the fitness evaluation term remains the primary computational bottleneck.

The total runtimes reported in Table~\ref{tab:runtime_decomposition} correspond to the optimization of one representative spatial \gls{MS} distribution on the reference 32 GB workstation.
The required number of equivalent 32 GB workstation nodes for an MS distribution update period $\Delta t$ is approximated as $N_{\mathrm{eq,CGA}} = T_{\mathrm{CGA}}/\Delta t$ and $N_{\mathrm{eq,PDGA}} = T_{\mathrm{PDGA}}/\Delta t$ for the CGA- and PDGA-based approaches, respectively.
These estimates reflect that independent optimization tasks, such as separate evolutionary runs, are executed in parallel across equivalent computational nodes.
Table~\ref{tab:hardware_requirements} summarizes the resulting hardware requirements for different update periods.
The results indicate that the proposed framework is not intended for minute-level re-optimization, as such update rates would require substantially increased computational resources, falling outside the intended operational regime.
In contrast, when spatial \gls{MS} distributions evolve over hourly or multi-hour timescales, the computational requirements become moderate and practically attainable.
In particular, the \gls{PDGA}-based approach requires few equivalent workstation nodes for update periods on the order of hours, making it suitable for offline and even semi-online optimization on small-scale workstation or server infrastructures.
Conversely, the \gls{CGA}-based approach demands substantially more computational resources for the same update periods, although it remains affordable under parallel execution.

\section{CONCLUSION}
\label{sec:Conclusion}

This paper has presented two evolutionary \gls{AP} sleep-mode strategies, \gls{CGA}-\gls{ASW} and \gls{PDGA}-\gls{ASW}, to improve the energy efficiency of large-scale \gls{CF-mMIMO} networks under spatially heterogeneous traffic. 
Both methods exploit genetic-algorithm principles to explore the activation space more effectively than greedy heuristics, achieving near-optimal ON/OFF configurations through population diversity and stochastic search.
Results show consistent and significant gains over a state-of-the-art greedy baseline in both \gls{EE} and \gls{SE}.
In particular, \gls{PDGA}-\gls{ASW} leverages multi-objective search to characterize the \gls{EE}–\gls{SE} trade-off via the Pareto front, whereas \gls{CGA}-\gls{ASW}, combined with a discrete interval-refinement or bisection procedure, identifies the \gls{AP} configuration that maximizes network \gls{EE}.
Overall, evolutionary optimization provides a robust and practical solution for \gls{AP} activation, especially in highly inhomogeneous scenarios with exponentially growing search spaces, achieving performance close to the best solutions observed via extensive Monte Carlo evaluations.

The computational feasibility of the proposed framework has also been analyzed, including runtime and hardware requirements.
Results indicate that \gls{GA}-based optimization is feasible for spatial \glspl{MS} distributions evolving over slow time scales (e.g., hours), enabling offline or semi-online operation.
\gls{MMSE} processing incurs higher runtime than \gls{CB} due to its enhanced interference suppression, while GA-based methods require longer runtimes than the GoF-\gls{ASW} baseline in exchange for improved \gls{EE} performance.
\gls{PDGA}-\gls{ASW} requires fewer fitness evaluations and achieves lower runtime than \gls{CGA}-\gls{ASW}, highlighting its superior computational efficiency.
Runtime analysis shows that fitness-function evaluations dominate the overall cost, far exceeding auxiliary operations.
These findings support the practical feasibility of the framework using small-scale workstations or servers, particularly for network planning scenarios with slowly varying spatial traffic distributions.

The proposed framework is deployment-agnostic, enabling its application across different spatial \gls{MS} distributions, traffic loads, and \gls{AP} densities without requiring structural modifications.
Future work will address the analysis of highly loaded large-scale networks, as well as the incorporation of fronthaul limitations.
Additional directions include hybrid approaches that combine data-driven prediction with evolutionary optimization, to further assess the applicability of the proposed methods in more demanding scenarios.


\bibliographystyle{IEEEtran}
\bibliography{Cell_free_biblio}

\end{document}